%% file: main.tex
\documentclass[conference]{IEEEtran}

\title{EmPoWeb: Empowering Web Applications with Browser Extensions}

\author{\IEEEauthorblockN{Doli\`ere Francis Som\'e} \\
\IEEEauthorblockA{\textit{Universit\'e C\^ote d'Azur / Inria, France} \\
doliere.some@inria.fr}
}

\usepackage{graphicx}
\usepackage{listings}
\usepackage{color}
\usepackage{caption}
\usepackage{footnote}
\usepackage{amsmath}
\usepackage{amssymb}
\usepackage{longtable}
\usepackage[table]{xcolor}
\usepackage{fontawesome}
\usepackage[bookmarks=false]{hyperref}

\input{macros}

\begin{document}
\maketitle

\begin{abstract}
\input{fabstract}
\end{abstract}

\input{fintro}
\input{fback}

\input{fmethod}
\input{tool}

\input{fres}
\input{casestudy}

\input{ndiscuss}

\input{related_work}
\input{conclusion}

\section*{Acknowledgments}
We would like to thank the anonymous reviewers and the PC for the quality of the comments and revision expectations as they helped us improve the work. Special thanks to Nadege Som\'e, Sadry Fievet, Nataliia Bielova, and Tamara Rezk  for proof-reading, support and insightful discussions, comments and suggestions.
The author is grateful to Inria Sophia Antipolis - M\'editerran\'ee "Nef" computation cluster for providing resources and support. 
This work was supported by the project PIA ANSWER.



\bibliographystyle{IEEEtran}
\bibliography{IEEEabrv,main}

\input{extensions}

\end{document}

%% file: macros.tex
\def\code#1{\texttt {{#1}}}

\def\manifest{\code{manifest.json}}
\def\permissions{\code{permissions}}
\def\host{\code{host}}
\def\eval{\code{eval}}

\def\storage{\code{storage}}
\def\bookmarks{\code{bookmarks}}
\def\cookies{\code{cookies}}
\def\history{\code{history}}
\def\topsites{\code{topSites}}
\def\downloads{\code{downloads}}

\def\management{\code{management}}
\def\externally{\code{externally\_connectable}}

\def\tabs{\code{tabs}}

\usepackage{graphicx}
\usepackage{listings}
\usepackage{hyperref}
\usepackage{color}
\usepackage{caption}
\usepackage{footnote}
\usepackage{amsmath}
\usepackage{amssymb}
\usepackage{longtable}
\usepackage[table]{xcolor}
\usepackage{array}
\newcolumntype{P}[1]{>{\centering\arraybackslash}p{#1}}
\definecolor{LightCyan}{rgb}{0.88,1,1}
\makesavenoteenv{tabular}

\makeatletter
\def\footnoterule{\kern-3\p@
  \hrule \@width 2in \kern 2.6\p@} 
\makeatother

\definecolor{lightgray}{rgb}{.9,.9,.9}
\definecolor{darkgray}{rgb}{.4,.4,.4}
\definecolor{purple}{rgb}{0.65, 0.12, 0.82}

\lstdefinelanguage{JavaScript}{
  keywords={typeof, new, true, false, catch, function, return, null, catch, switch, var, if, for, in, while, do, else, case, break},
  keywordstyle=\color{blue}\bfseries,
  ndkeywords={document, window, chrome, addEventListener, postMessage, get, set, storage, addListener, onMessageExternal, class, export, boolean, throw, implements, import, this},
  ndkeywordstyle=\color{darkgray}\bfseries,
  identifierstyle=\color{black},
  sensitive=false,
  comment=[l]{//},
  morecomment=[s]{/*}{*/},
  commentstyle=\color{purple}\ttfamily,
  stringstyle=\color{red}\ttfamily,
  morestring=[b]',
  morestring=[b]"
}

\lstdefinelanguage{CSP}{
  keywords={default-src, script-src, child-src, frame-src, script-src, style-src, report-uri, connect-src, img-src, object-src, frame-ancestors, plugin-types, form-action, sandbox, worker-src, font-src, media-src},
  ndkeywords={content\_security\_policy, 'self', 'unsafe-eval'},
  morestring=[b]',
  alsoletter={:},
  alsodigit={-}
}

%% file: fabstract.tex
Browser extensions are third party programs, tightly integrated to browsers, where they execute with elevated privileges in order to provide users with additional functionalities. Unlike web applications, extensions are not subject to the Same Origin Policy (SOP) and therefore can read and write user data on any web application. They also have access to sensitive user information including browsing history, bookmarks, credentials (cookies) and list of installed extensions. They have access to a permanent storage in which they can store data as long as they are installed in the user's browser. They can trigger the download of arbitrary files and save them on the user's device.

For security reasons, browser extensions and web applications are executed in separate contexts. Nonetheless, in all major browsers, extensions and web applications can interact by exchanging messages. 
Through these communication channels, a web application can exploit extension privileged capabilities and thereby access and exfiltrate sensitive user information.

In this work, we 
analyzed the communication interfaces exposed to web applications by Chrome, Firefox and Opera browser extensions.
As a result, we identified many extensions that web applications can exploit to access privileged capabilities. Through extensions' APIS, web applications can bypass SOP and access user data on any other web application, access user credentials (cookies), browsing history, bookmarks, list of installed extensions, extensions storage, and download and save arbitrary files in the user's device.

Our results demonstrate that the communications between browser extensions and web applications pose serious security and privacy threats to browsers, web applications and more importantly to users. We discuss countermeasures and proposals, and believe that our study and in particular the tool we used to detect and exploit these threats, can be used as part of extensions review process by browser vendors to help them identify and fix the aforementioned problems in extensions. 


%% file: fintro.tex
\section{Introduction}
Browser extensions or addons are third party programs, that can extend the functionality of browsers and improve users' browsing experience. 
In this work, we focus on the WebExtensions API, a cross-browser extensions system compatible with major browsers including including Chrome, Firefox, Opera and Microsoft Edge~\cite{ChromeExtensionsAPI, FirefoxWebExtensionsAPI, OperaExtensionsAPI, MicrosoftEdgeExtensionsAPI}. 
Tightly integrated to browsers, extensions have access to elevated browser APIs and features. For instance, they can make HTTP requests to get data from any web application server, including those where users are logged into, such as their mailing, banking, social network applications, etc. As a comparison, web applications are bound by the Same Origin Policy (SOP)~\cite{SOP} and cannot access other web applications data, unless they both implement mechanisms such as Cross-Origin Resource Sharing (CORS)~\cite{CORS}.


Due to their privileged position in browsers, it is well understood that extensions pose serious security and privacy threats to user data~\cite{Louw-etal-07-DIMVA, Bart-etal-10-NDSS, Band-etal-10-USENIX, Carl-etal-12-USENIX, Onar-etal-13-DIMVA, Onar-etal-15-CS, Calz-etal-15-ESOP}. 
Therefore, in order to limit extensions capabilities, a mandatory permission system requires that extensions explicitly declare the set of APIs they effectively need to access. Nonetheless studies have shown that extensions still require many permissions~\cite{Guha-etal-11-SP, Kapra-etal-14-USENIX, Heul-etal-15-HOTOS}. Extensions also go through a review process from browser vendors. But again studies have unveiled many malicious extensions circumventing the review process to exfiltrate sensitive user data~\cite{Kapra-etal-14-USENIX, Weis-etal-17-CSAC, Star-Niki-17-WWW}

Besides, a benign (non-malicious) extension can be buggy, allowing adversaries to exploit its vulnerabilities in order to get access to user sensitive data.  
One type of adversary that can exploit such vulnerabilities in extensions is the \code{web attacker}~\cite{Band-etal-10-USENIX, Carl-etal-12-USENIX, Calz-etal-15-ESOP}. 
Indeed for security reasons, extensions and web applications execute in different and isolated contexts. Nonetheless, extensions have access to the DOM of webpages. Extensions and webpages can also set up communication channels to exchange data with one another using the \code{postMessage} API~\cite{CrossOriginCommunications} for instance.
Bandhakavi et al.~\cite{Band-etal-10-USENIX} and Carlini et al.~\cite{Carl-etal-12-USENIX} found that a few extensions manipulate data extracted from webpages without sanitization, leading to privileged escalations, because such data can be influenced by a \code{web attacker}. 
Calzavara et al.~\cite{Calz-etal-15-ESOP} found that message passing APIs can be abused for privilege escalation, and formally characterized the privileges that be exploited by a \code{web attacker}.

Similarly to those studies, our threat model considers the web application as the attacker. Specifically, we seek to study at large scale, the security and privacy implications of message passing APIs~\cite{Calz-etal-15-ESOP} among extensions in the wild,  whether there are extensions that can be exploited by web applications to access sensitive user information. 
For instance, an extension can set up an interface (register a listener) to receive from web applications, messages consisting of the URL of a resource (data) hosted by another web application. The extension then makes a request to fetch the data (since it can do so with any web application as it is not subject to the SOP) and returns the response to the web application that sent the URL. 
Hence, these communications channels are a way for a deliberately or accidentally vulnerable extension to indirectly give a web application access to browser features and APIs that the web application is not directly allowed to access.

%

We built a static analyzer and applied it to the message passing interfaces exposed by Google Chrome, Firefox and Opera extensions to web applications. When the tool found that a privileged extension capability could potentially be exploited by web applications, the extension was flagged suspicious.
By manually reviewing the code of suspicious extensions, we found that 197 of them (mostly on Chrome) can be exploited by web applications (attackers) to access elevated browser features and APIs and sensitive user information. The extensions we have found have vulnerabilities that can be exploited by web applications to (i) break the privilege separation between extensions and web applications and execute arbitrary code in extensions context, (ii) bypass the Same Origin Policy and access user data on other applications, (iii) read user cookies and use them to mount session hijacking attacks~\cite{SessionHijacking}, (iv) access data such as user browsing history, bookmarks, list of installed extensions that besides violating user privacy can be used for tracking purposes~\cite{Sjos-etal-17-CODASPY, Star-Niki-17-SP, Sanc-etal-17-USENIX, Guly-etal-18-WPES}, (v) store and retrieve data from extensions persistent storage for tracking purposes and (vi) trigger the download of malicious software on the user device which execution can damage user data.

In summary, this paper shows the security and privacy threats associated with the interactions between browser extensions and web applications and makes the following contributions:

\begin{itemize}
	\item We built a static analysis tool and analyzed extensions message passing interfaces at large-scale: 66,401, 9,391 and 2,523 extensions on Chrome, Firefox and Opera respectively. 
	About 4.97\%, 5.14\% and 8.48\% of Chrome, Firefox and Opera extensions respectively were flagged as suspicious.
	\item We identified 197 extensions that pose various security and privacy threats to browsers, web applications, and users. They can be exploited by web applications to bypass the SOP, read user cookies, browsing history, bookmarks, list of installed extensions, store and retrieve data from the extension storage, or download malicious files and store them on the user device. 
\end{itemize}
We provide \href{http://www-sop.inria.fr/members/Doliere.Some/empoweb/}{online a web page} for navigating through the different results of this work, including the \href{http://www-sop.inria.fr/members/Doliere.Some/empoweb/extsanalyzer/}{tool} and \href{http://www-sop.inria.fr/members/Doliere.Some/empoweb/extensions/}{videos} demonstrating how we exploited the capabilities of some of the extensions.

In the beginning of October 2018, we reported our findings to the vendors who positively acknowledged the issues. For instance, Firefox has already removed all the reported extensions, and Opera all but 2. We are still discussing with Opera on how to fix the 2 extensions. With Chrome, we are also discussing with them on the actions to take. 
%
%
That notwithstanding, we argue that browser vendors need to review extensions more rigorously, in particular take into consideration the use of message passing interfaces in extensions. We think that tools such our static analyzer can help in identifying and fixing the security and privacy threats we have identified in this paper. 
We also discuss a few proposals on  the current design of the message passing interfaces so as to mitigate the attacks that web applications can mount against extensions. 

%% file: fback.tex
\section{Background}
\label{sec:background}

\subsection{Browser extensions capabilities}

Standard web technologies (HTML, CSS and JavaScript) are used to develop extensions for major web browsers including Chrome, Opera, Firefox, and Microsoft Edge. Interestingly, their specific extensions APIs,~\cite{ChromeExtensionsAPI, OperaExtensionsAPI, FirefoxWebExtensionsAPI, MicrosoftEdgeExtensionsAPI} are compatible with each other to some extent, making it easy to migrate extensions written for a specific browser to other browsers with just a few changes. This work focuses on these extensions, referred to as WebExtensions.

{\bf{Extensions security considerations}}
Extensions are third party software, that users install to alter their browsers behavior and improve their browsing experience. Tightly integrated to browsers, extensions have access to privileged browser features,  making them interesting targets for attackers. Hence, to limit the harm that attackers could cause if they take control of an extension, the APIs that extensions effectively have access to  are only those for which they have explicitly requested the related permission in their \manifest\ file. Listing~\ref{lst:manifest_file} shows an example of manifest file and the \permissions\ (features and APIs) the extension will be granted at runtime.

\begin{lstlisting}[language=JavaScript,caption={Permissions declaration in a manifest file}, label={lst:manifest_file}]
{
    "permissions": [
      "<all_urls>", 
      "storage", 
      "management", 
      "cookies"
      "history", 
      "bookmarks", 
      "downloads",
    ]
}
\end{lstlisting}
These are only a subset of all the capabilities provided by browsers to extensions.
When installed, this extension will be granted full access (read/write) to data on any web application, thanks to its \host\ permission \code{<all\_urls>}. This implies that if the user is logged into a web application (such as mailing, banking, social networks), the extension also has access to the user's private data on that application. 
The rest of the permissions read straightforwardly. The \storage\ permission allows the extension to store and retrieve data in the browser as long as it is installed. The permissions \management, \cookies, \history, and \bookmarks\ give the extension access to the list of installed extensions, web applications cookies, user browsing history and bookmarks respectively. With the \downloads\ permission, the extension can download and save arbitrary files in the user device. 
At runtime, those APIs (and any extension-specific API in general) are all accessible via the \code{chrome} object in Chrome and Opera browsers, and via \code{browser} object in Firefox and Microsoft Edge. To ease the readability of this work, we often omit the \code{chrome} and \code{browser} from extensions APIs names.

{\bf{Architecture}}

\begin{figure}[ht]
\includegraphics[width=0.5\textwidth]{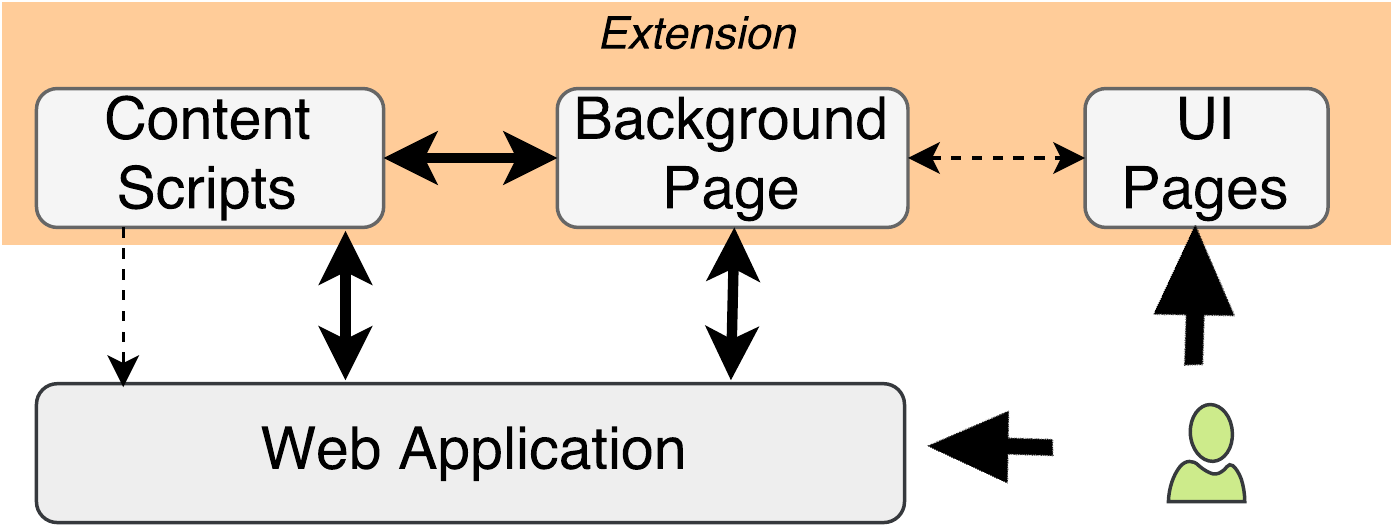}
\captionsetup{justification=centering}
\caption{Browser extensions architecture - Communications with web applications}
\label{fig:extension_architecture}
\end{figure}

Extensions can be divided in three main parts, as shown in Figure~\ref{fig:extension_architecture}. 
The background page is the main part of the extension. It has full access to all the capabilities of the extension. Users interact with the extension through UI pages (i.e. UI elements, options and setting pages), in order to enable or disable it, or customize its behavior. UI pages also have access to the full capabilities of the extension. Content scripts are injected by extensions to run along web applications. Even though they are not granted access to all the extension capabilities, they can directly use the \host\ and \storage\ permissions to access user data on any web application or to store and retrieve data from the extension storage. Content scripts can also manipulate the DOM of webpages~\cite{DOM} and inject content in them. On Chrome and Opera, each extension is assigned a permanent unique identifier, which is the same for all users of the extension. Firefox however generates a random identifier for the extension, per user browser~\cite{FirefoxUUID}.

\subsection{Interactions}
\label{sec:coms_interactions}
Background and UI pages have direct access to each other's execution contexts~\cite{BrowsingContext}, but content scripts execute in a separate context.
Web applications run in yet other separate execution contexts. Nonetheless, content scripts have direct access to web applications localStorage, DOM, and execution context, where they can inject and execute arbitrary scripts.

Even though content scripts, background pages and web applications run in separate execution contexts, they can establish communication channels to exchange messages  with one another~\cite{ChromeMessaging, OperaMessaging} as shown in Figure~\ref{fig:extension_architecture}.
We describe below 
the APIs for sending and receiving (listening for) messages between the content scripts, background pages and web applications.

{\bf{Content scripts and background pages}}
There are two types of communication channels: one-time and long-lived channels. One-time channels are opened to send a message and are closed after the response is received. Long-lived channels, connections or ports, are maintained open to exchange multiple messages. A port can have a name in order to distinguish it from other long-lived channels. 

For one-time messages, content scripts use the \code{runtime.sendMessage} API to send messages to background pages. Similarly, background pages employ the \code{tabs.sendMessage} API to send messages to content scripts. For receiving messages, both components can invoke the \code{runtime.onMessage.addListener} API.

Similarly, \code{runtime.onConnect.addListener} and \code{runtime.connect} are used to  establish long-term communications between background pages and content scripts.

{\bf{Web applications and content scripts}}
Exchanges between web applications and content scripts are achieved with the Cross-Origin Communications API~\cite{CrossOriginCommunications}: \code{postMessage} is used for sending messages, and \code{onmessage} or \code{addEventListener} to receive messages. Below is a listing which shows how messages are sent and received between web applications and content scripts. 

\begin{lstlisting}
  // Send and receive
postMessage("Hello Extension", "*");
addEventListener("message", function(event){
  Received_response = event.data;
});
  // Receive and Reply
addEventListener("message", function(event){ 
  Received_message = event.data;
  postMessage("Hello Web Application", "*")
});
\end{lstlisting}

In this example, the web application sends the message \emph{Hello Extension} to the content script, which receives and writes it in the variable \code{Received\_message}. Then it replies with \emph{Hello Web application}, which the web application receives and saves in the variable \code{Received\_response}.

{\bf{Web applications and background pages}}
On Chrome and Opera, web applications can also directly communicate with extensions background pages. To do so, extensions have to declare in their \manifest\ file, using the \externally\ key, the list of web applications, where communication with  the background page is allowed. For security reasons, one cannot use wildcard (for instance *) to allow communications between the background pages and all web applications. Additionally, communications can only be initiated by web applications. 

The \code{runtime.sendMessage} and \code{runtime.connect} APIs are exposed to web applications in Chrome and Opera, and can be used to send one-time messages or establish long-term connections with background pages. Conversely, the APIs \code{runtime.onMessageExternal.addListener} and \code{runtime.onConnectExternal.addListener} can be used in the background page, to receive and reply to messages sent by web applications.
Below is an example of how to send a message from a web application to the background page of an extension which unique identifier is \code{ExtensionID}. 
\begin{lstlisting}
  // Send and Receive
chrome.runtime.sendMessage(ExtensionID, "Hello Extension", function(response){
  Received_response = response;
});
  // Recieve and Reply
chrome.runtime.onMessageExternal.addListener(function(message, sender, sendResponse){
  Received_message = message;
  sendResponse("Hello Web application");
})
\end{lstlisting}
The application sends \emph{Hello Extension} to the background page which replies with \emph{Hello Web application}.

\subsection{Threat models}
\label{sec:threats}
An attacker is a script that is present in a web application currently running in the user browser. The script either belongs to the web application or to a third party. 
The goal of the attacker is to interact with installed extensions, in order to access user sensitive information. It relies on extensions whose privileged capabilities can be exploited via an exchange of messages with scripts in the web application.
We consider the following security and privacy threats posed by extensions.
\begin{enumerate}
  \item {\bf{Execute code}}: these are extensions that can be exploited by the attacker to execute arbitrary codes in the extension context. Executing code in the background page gives the attacker access to all the capabilities of the extension. In content scripts, the attacker can bypass SOP by making cross-origin AJAX requests, and use the extension permanent storage for tracking purposes.
  \item {\bf{Bypass SOP}}: in this case, an attacker can exploit the capability of the extension to make cross-origin requests without being restricted by the Same Origin Policy. 
  \item {\bf{Read Cookies}}: the attacker can read the user cookies and use them to mount session hijacking attacks, access user data and take actions on her behalf.
  \item {\bf{Trigger Downloads}}: the attacker exploits extensions to trigger the download of arbitrary malicious files (software) and saves them on the user's device without requiring any action from the user. If the user inadvertently runs such software, the attacker takes control of her device and performs malicious actions.
  \item {\bf{Read browsing history, bookmarks and list of installed extensions}}: these information reveal the user interests and habits and can be used by the attacker for tracking purposes, or to serve targeted and personalized advertisement.
  \item {\bf{Store data}}: the attacker can store and retrieve information in the extension storage. This can be used for tracking purposes, even though users clear web applications storages.
\end{enumerate}
For the sake of simplicity, we often refer to the attacker as the web application in which it runs.

%% file: fmethod.tex
\section{Methodology}
\begin{figure}[ht]
\includegraphics[width=0.5\textwidth]{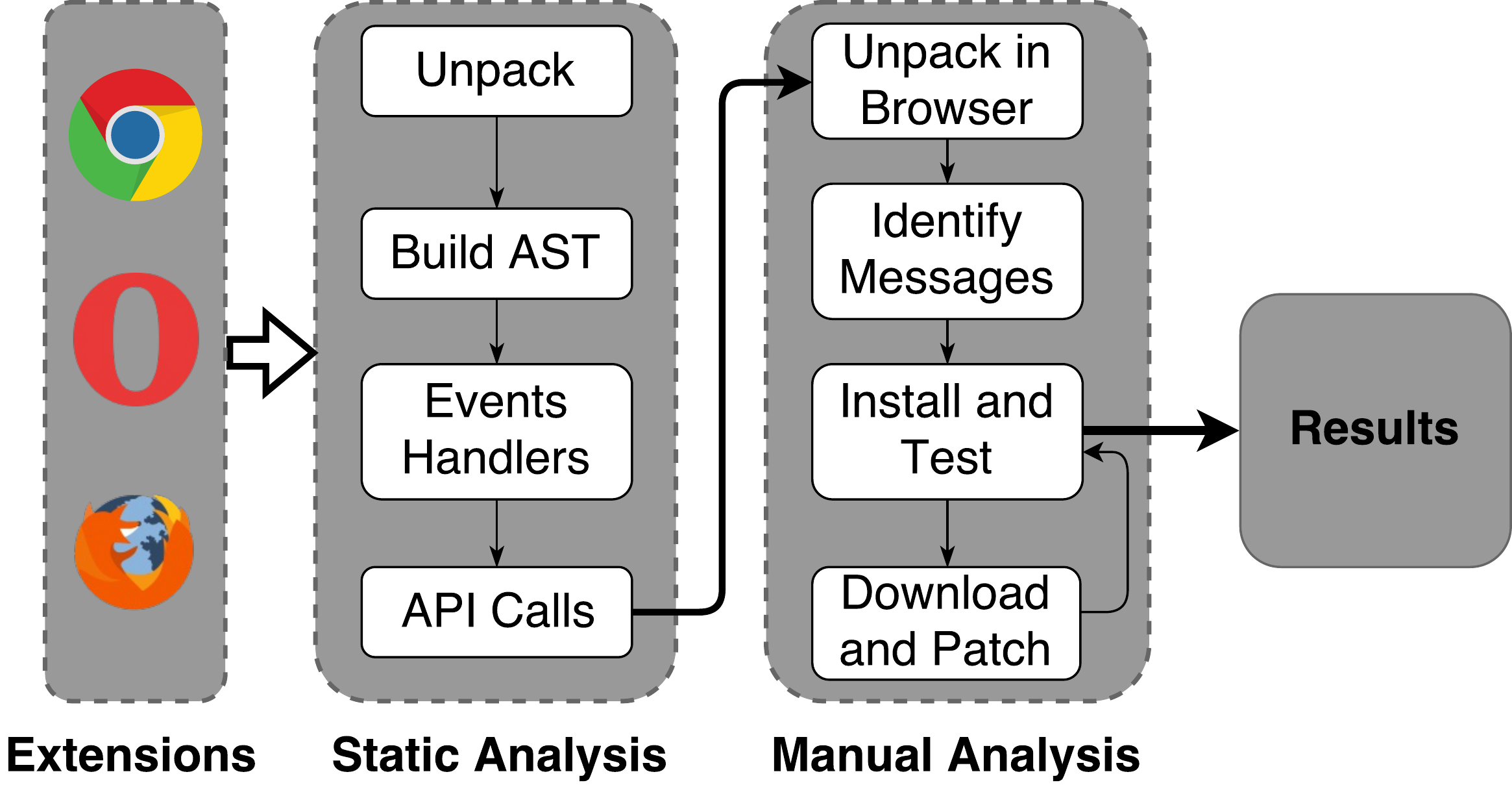}
\captionsetup{justification=centering}
\caption{Methodology - static and manual analysis}
\label{fig:methodology}
\end{figure}

We built a static analyzer that detects suspicious communications enabled by extensions with web applications.  
To identify extensions that are potentially concerned with the security and privacy threats identified in the previous section, we focus on 78,315 extensions from Chrome, Firefox and Opera browsers. 
Then we manually reviewed the code of the extensions to precisely validate the results of the static analyzer, and more importantly to construct the signatures of the messages that have to be exchanged with extensions to successfully exploit their capabilities. 
Figure~\ref{fig:methodology} shows the analysis process.

\subsection{Static analysis}
The goal of the static analyzer is to report only extensions that potentially pose a security and privacy threat, in order to reduce false positives as much as possible, and reduce the burden of the manual analysis. It has been fully written in JavaScript, using various Node.js packages. We used Esprima~\cite{Esprima} and Recast~\cite{Recast}, for parsing and manipulating JavaScript abstract syntax trees (AST), and Jsdom~\cite{Jsdom} for parsing HTML.

{\bf{Unpack extensions and gather scripts}}
We crawled extensions using SlimerJS Browser Automation tool~\cite{SlimerJS}.
In the extension \manifest\ file, background pages are either declared by a set of scripts files, or an HTML file, which further includes the scripts of the background page. UI pages are built as HTML pages, and also indicated in \manifest\ file. The Jsdom HTML parser served here to extract scripts embedded in background as well as UI pages.  
Static content scripts are directly declared in the \manifest\ file. 
Background and UI pages can further dynamically inject content scripts in web applications, by calling the \code{tabs.executeScript} API. Those were also extracted by analyzing the AST of background and UI pages scripts, and analyzed as other content scripts.

{\bf{Parse scripts and build AST}}
Scripts were parsed with Esprima, resulting in an AST~\cite{AST}, which contains all JavaScript constructs used in content scripts, background and UI pages scripts. Almost everything in JavaScript is an object~\cite{ECMASCRIPT}. To ease manipulation of the AST, the following additional actions were taken to build three indexed tables of assignments to variables and object properties (\code{assignments}), function definitions/expressions and object methods (\code{functions}), and finally functions and object methods invocations (\code{calls}). Basically, those are key/value pairs, in which the keys in the tables corresponded to the names of variables, object properties and functions.
Each entry was then associated with a list of all possible values it could resolve to. For assignments, the values were all expressions assigned to a variable or object. For function definitions and object methods, the values were the parameters and body of the function. Finally, for function calls, the values associated to their names in the indexed table were their invocation arguments. The static analyzer successfully handled functions defined using the \code{bind} method, and functions invoked using the \code{call} or \code{apply} methods.

{\bf{Event handlers of page messages APIs}}
For each message listener (See Section~\ref{sec:background}) in content scripts, background and UI pages, we first looked up the indexed table of function invocations (\code{calls}) to search whether the extension registered listeners for messages from the web applications (a call to \code{addEventListener} API for instance in content scripts). 
In browser contexts, all JavaScript objects are properties of a global object named \code{window}.
Different aliases, \code{this, self, global}, are sometimes used to refer to the \code{window} object~\cite{WindowObject, JavaScriptScope}. JavaScript object properties can be accessed using the dot and the array or bracket notations~\cite{JavaScriptPropertyAccess}. For the sake of simplicity, we considered the dot notation and the bracket notation when the property name was a literal (a string).
Considering the global object names (\code{window, top, self, this}), and JavaScript dot and bracket property accesses, we generated the different ways an API can be invoked. 
For instance, \code{addEventListener} can be called in 9 different ways \code{addEventListener, window.addEventListener, window["addEventListener"]} and others. In general, we consider that an object could be accessed in 9 different ways, its properties in 18 different ways, the properties of its properties in 36 ways and so forth. 
When we found an invocation to communications APIs in content scripts, background and UI pages, we extracted their arguments and resolved them as follows.

For \code{addEventListener}, the first argument should be the literal \code{message}, and the second argument a function. Otherwise, we use the indexed table of \code{assignments} and \code{functions} to resolve them to the literal \code{message} and a function respectively. 
Resolving an argument simply consist in checking whether the indexed table has an entry which key matches the argument name, and further checking whether any of its associated values resolve to the type and value we expect the argument to have. For \code{addEventListener}, we expect the first argument to be a \code{Literal} and have the value \code{message}. Its second argument is expected to be a function. We follow the same process to extract all message handlers (listeners) in content scripts, background and UI pages.


{\bf{Sensitive APIs Calls}}
The handlers (functions) of web applications messages in extensions are parsed to extract all their constructs. If the handlers further call other functions, those functions are looked up using the indexed table, and their bodies parsed to also extract their constructs. Finally, the constructs are analyzed to decide whether the extension potentially poses any of the security and privacy threats considered in this work.
\begin{itemize}
	\item An extension is flagged as potentially executing arbitrary code sent from web applications if it invokes functions like \code{eval} (in any part of the extension) or \code{tabs.executeScript} (in background and UI pages).
	\item An extension is flagged as potentially allowing web applications to bypass SOP, if its constructs include APIs that can be used to make AJAX calls. This includes the creation of new \code{XMLHttpRequest} objects, calls to \code{fetch} API, or any AJAX specific API provided by popular third party libraries such jQuery and AngularJS (\code{\$.get, \$.ajax, \$.post, \$http.get, \$http.post}).
	\item If the constructs include calls to \storage\ API such as \code{storage.local.set, storage.local.get, storage.sync.set, storage.sync.get}, then the extension is flagged as potentially storing/retrieving data for web applications.
	\item An extension is considered as potentially leaking user cookies, history, bookmarks, and list of extensions to web applications if either of the following invocations were found in their message handlers constructs: \code{cookies.getAll, history.search, history.getVisits, bookmarks.getTree, management.getAll}, and related APIs.
	\item Finally, an extension is considered as probably allowing web applications to download and save files in the user computer (device) if their messages event handlers constructs include invocation to \code{downloads.download}.
\end{itemize}
It is worth mentioning the case of content scripts forwarding messages to background pages. When this is the case, the constructs of content scripts messages handlers in the background pages are also analyzed, looking for calls to any API which potentially poses security and privacy threats. In fact, content scripts only have access to the \host\ and \storage\ capabilities. When they need access to more capabilities, they can send messages to the background pages which may then give them access to the related capability. Content scripts can forward messages they receive from web applications, to the background page. The latter handles the message and responds to the content scripts which in turn respond to the application. This is particularly true in Firefox which does not allow direct communications between web applications and background pages. Nonetheless, we have observed many content scripts forwarding messages to background pages, even to access APIs they can directly use from their own context.

\subsection{Manual Analysis}
Recall that for each suspicious extension, the tool reports precisely (i) the component(s) (content scripts, background or UI pages) and the file or set of files to analyse, (iii) the name of the message passing interfaces registered (iv) the handlers of the message passing interfaces and other functions that are called from those handlers (v) and finally, the sensitive APIs that are triggered in the handler and its called functions. 
So the goal of the manual analysis was to confirm the suspicion of the static analyzer, build and test the precise signatures of messages that had to be sent by web applications to exploit extensions capabilities. Following is how we typically manually review an extension. 

{\bf{Unpack in Browser}}
First, we download and unpack the extension code directly in a browser, using the CRX Extension Viewer~\cite{CRXViewer}\footnote{CRX Extension Viewer~\cite{CRXViewer} is an handy extension that can download and nicely display in the browser the content of extensions, allowing to navigate their files very conveniently}. We locate the files to be analyzed, according to the concerned component(s). In the particular case of content scripts, the tool reports the precise index in the content scripts array containing the file or set of files that has to be analyzed~\footnote{Indeed, content scripts are declared as an array of a set of files, so the tool reports the index of the files to analyse}. 

{\bf{Identify Messages}} This step is concerned with building the payloads or signature of messages that can be sent to the extension to exploit is capabilities. 
In each file, we search for the message passing interfaces, their handlers and the functions they invoke. In those handlers and functions, we look for the sensitive APIs that are triggered and more importantly, the parts of the received messages that trigger calls to the extensions sensitive APIs. To build the payloads, 
we carefully inspect how objects received as parameters in the message passing interfaces handlers (and related functions) are manipulated, which properties of the objects are accessed, the names of the properties, and their values. This gives us the precise signature of the messages. Two situations arise here. Either we found that the signature of the messages are predefined by the extension, in which case they cannot be influenced by the attacker (See Section~\ref{sec:otherthreats} for more details), and we stop the analysis and consider the extension as a false positive. Otherwise, we continue the analysis by installing and interacting with the extension by sending messages according to the signature that we have found.

{\bf{Install and Test -- Download and Patch}} The final step consists in mounting the exploits against the vulnerable extension with the payloads built in the previous step. 
Those interactions are done from the Browser console~\cite{BrowserConsole}, after we navigate to the websites in which the extension injects its files. Sometimes, those websites require users to be logged in. We would create accounts and navigate to them. In rare cases, we could not create accounts or the websites were not responding. So we download the extension, and modify its permissions in order to make it inject files in websites that are accessible (i.e. \url{http://localhost/}), from were  we mount the attack.
During the tests, we use the browser debugger, set breakpoints in the extension codes in order to track the propagation of messages and calls in the extension code.



{\bf{Time taken to manually analyze extensions}}
It is rather difficult to precisely evaluate how much time it takes to manually analyze an extension. Indeed, this work went through 3 phases. We did a first crawl of extensions in the middle of November 2017 and run our static analyzer to test it.  It took around 2 months and half ( till the beginning of February 2018)  to come with a mature analyzer. But during that phase, we had already discovered almost 87\% of the extensions reported in this paper. 
Then, from the beginning of February 2018 we analyzed all the extensions again in around 4 weeks. 
Finally, in mid of May 2018, we did a new crawl and analysis, and it took around 10  days to vet the extensions. 
Again, for most of the extensions, we had already tested them, built the signatures of the messages, so analyzing them again consisted only in checking that they were still exploitable. 

So overall, what we observed is that during the phase we implemented and tested the static analyzer, manually reviewing the code of a suspicious extension was long, because the process was rather imprecise and not straightforward. Then throughout that period of tests, we had acquired a lot of expertise in the review process, which made it faster in the end. For instance, we have started to recognize similar patterns and codes in extensions (many extensions reuse similar code) and therefore we knew when they had to be skipped or not an extension. 
Currently, we think that for an expert,
15 mn would be a sufficient average time to correctly review an extension for the threats that we have reported. In practice, some extensions will take longer to analyse while others will be analyzed in a couple of minutes. 

\subsection{Limitations and Challenges}
First note that we considered only scripts that are part of the extension packages. For instance, background and UI pages may reference external scripts. Those scripts were not considered in our analysis. Nonetheless, we think that extensions bundles are more likely to contain most of the APIs that we consider in this work, as extensions developers are recommended to avoid referencing remote scripts in extensions codes.

Our static analysis tool suffers from many limitations. The first one is the fact that we did not consider scopes~\cite{JavaScriptScope}, which lead to unnecessary functions being analyzed. However, this is not a problem ultimately because all the results were further manually reviewed to remove false positives. It also suffers from some false negatives, mainly because of the flexibility of JavaScript that make it challenging to exhaustively address all the ways message listeners can be invoked in extensions. 

Manually analyzing complex, large and sometimes minified and obfuscated JavaScript programs making use of callbacks everywhere is not trivial for a single person. But we have taken all the time that was necessary to correctly perform the study. 
Finally, for a very few extensions, despite all our efforts at the static and manual analysis levels, we could not draw any conclusion about the potential threats they may pose. 


%% file: tool.tex
\section{Tool for analyzing communications APIs}
\label{sec:tool}
This section demonstrates a case study of the tool. We have released online a web version of the \href{http://www-sop.inria.fr/members/Doliere.Some/empoweb/extsanalyzer/}{tool}. It can be used to analyzed extensions directly in a browser. 

{\bf{Result of the static analyzer}}
Listing~\ref{lst:tool_result} shows the result produced by the tool when applied to the \emph{eRail.in} Chrome extension~\cite{eRail.in}.
\begin{lstlisting}[caption={Result of analyzing the \emph{eRail.in} extension}, label={lst:tool_result}]
{
    "com_via_cs": {
        "to_back": {
            "back": {
                "ajax": {
                    "$.get": "",
                    "$.post": "",
                    "$.ajax": "",
                    "XMLHttpRequest": ""
                },
                "cookies": {
                    "chrome.cookies.getAll": "",
                    "chrome.cookies.remove": "",
                    "cookies": ""
                }
            }
        }
    }
}
\end{lstlisting}

\begin{itemize}
    \item \code{com\_via\_cs} implies that webpages can communicate with the extension via the content scripts, by using the \code{postMessage} API. This extension has only 1 content script. When there are multiple content scripts, the tool analyzes each of them independently and produces results corresponding to each of them.
    \item \code{to\_back} indicates that the messages sent by webpages to the content script are forwarded to the extension background page. 
    \item The tool found that two sensitive APIs are reached in the background page: AJAX requests with calls to the jQuery AJAX APIs (\code{\$.get, \$.post, \$.ajax}) and access to cookies with invocation to the \code{chrome.cookies.getAll} and \code{chrome.cookies.remove} APIs.
\end{itemize}

The main goal of the tool is to raise awareness about the fact that an attacker may potentially get access to the extension's privileged APIs. One can then further review the code to validate or refute the results of the tool. 
For instance, after manually vetting the code of the \emph{eRail.in} extension, we effectively confirm that any webpage can access all user cookies and make AJAX request to any domain. See Section~\ref{sec:casestudy} for more details about examples of messages to be sent to extensions to benefit from their privileged capabilities.

{\bf{Releasing the tool}}
%
In addition to the full static analyzer used in this work, we have also prepared an online version for analyzing the message passing APIs of extensions. The only difference with the version used in this work is that it does not handle dynamically injected content scripts. This was done for simplicity reasons. That notwithstanding, in order to analyze dynamic content scripts, one can simply declare them in the extension manifest as static content scripts. Both versions of the tool will be publicly released. 

There is room for further improving the tool. For instance,  lessons can be learnt from the state-of-the-art on JavaScript static analysis tools in order to improve the extraction of messages passing listeners and tracking the escalation of extensions sensitive APIs. The set of threats considered in this work can also be extended further with state-of-the-art extensions threats in the literature. 

%% file: fres.tex
\section{Empirical Study}

We downloaded Chrome~\cite{ChromeWebStore}, Opera~\cite{OperaExtensionsCatalog}, and Firefox~\cite{FirefoxAddons} extensions by the end of November 2017. 
The extensions were statically analyzed in the beginning of February 2018 --- on a cluster of 200 nodes mainly because of storage limitations on our own devices. This was preceded by a long period of tests during which we improved the static analyzer, and fixed the list of security and privacy threats. In the middle of May 2018, we did another crawl and analysis. The results presented here are for this second dataset.

In this section, we first give an overview of the results, then we discuss in more details each threat and the report extensions where it was found. 

\subsection{Overview}
Table~\ref{tab:results_overview} presents the number of extensions we collected and analyzed.
Chrome provides the largest share of extensions, followed by Firefox and Opera. Recall that for Firefox, we are considering only extensions built using the new WebExtensions API~\cite{XPCOM}, and not those using the XPCOM/XUL API~\cite{FirefoxWebExtensionsAPI}. 

The static analysis tool reported 3,996 suspicious extensions that we manually vetted.
The results of the manual analysis are also shown in Table~\ref{tab:results_overview}. 
As with the share of extensions, Chrome had the largest share of extensions with threats.
In a total of 197 extensions, only 16 were found on Firefox, 10 on Opera, and the 171 others are Chrome extensions.  
Note that some single extensions pose more than one threat at a time. 
All the 197 extensions reported here effectively posed at least one or more of the security and privacy threats described in Section~\ref{sec:background}. During the manual analysis, we also identified the messages to be sent in order to exploit their capabilities. The full list of the extensions and the threats that they pose are given in  Table~\ref{tab:results_all} in the Appendix, because of page limitations.

\begin{table}[ht]
   \centering
   \caption{Data overview}
    \rowcolors{2}{gray!25}{white}
    \label{tab:results_overview}
   \begin{tabular}{|l|c|c|c|c|}
    \rowcolor{gray!50}
   \hline
   & Chrome & Firefox & Opera & Total \\ \hline
   Extensions analyzed & 66,401 & 9,391 & 2,523 & 78,315\\ 
   Suspicious extensions & 3,303 & 483 & 210 & 3,996 \\ 
   Execute Code & 15 & 2 & 2 & 19 \\
   Bypass SOP & 48 & 9 & 6 & 63 \\
   Read Cookies & 8 & - & - & 8 \\
   Read History & 40 & - & - & 40 \\ 
   Read Bookmarks & 37 & 1 & - & 38 \\
   Get Extensions Installed & 33 & - & - & 33 \\
   Store/Retrieve Data & 85 & 2 & 3 & 90 \\ 
   Trigger Downloads & 29 & 5 & 2 & 36 \\ \hline
   \bf{Total of unique extensions} & \bf{171} & \bf{16} & \bf{10} & \bf{197} \\ \hline
   \end{tabular}
\end{table}

\begin{figure}[ht]
\includegraphics[width=0.5\textwidth]{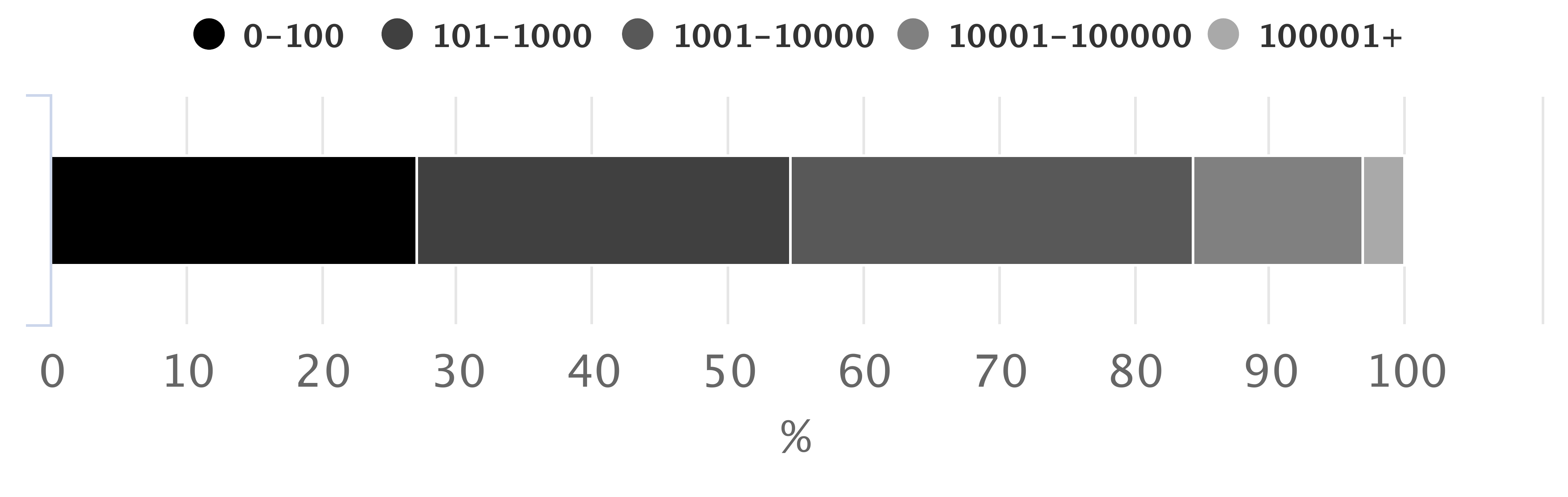}
\captionsetup{justification=centering}
\caption{Distribution of the number of users per extension}
\label{fig:extension_users}
\end{figure}

{\bf{Extensions installs and categories}}
Figure~\ref{fig:extension_users} presents the distribution of users impacted, or the number of installs per extension at the time of writing this paper. Around 55\% of the extensions have less than 1000 users, while the remainder 45\%  have thousands of installs, showing that those threats are present in rather popular extensions, hence affecting many users.
About 27\% of extensions have less than 100 users and another 27\% have between 100 and 1000 users. We see this as an opportunity for a tool such as ours to help improve extensions security, as it can serve to detect potentially malicious extensions while they are not yet very popular among users, thereby limiting their impact on users. 

Table~\ref{tab:extensions_category} further presents the category of these extensions. Note that the categorization of extensions is not done the same way by Chrome, Firefox and Opera browsers. Some categories exist only on specific browser, and not on others. Moreover, we found similar (or the exact same) extensions being differently classified depending on the browser. We tried to merge the different categories whenever possible. 

As one can observe, \code{Productivity} is the most popular category among the reported extensions. It is also the most popular category among all Chrome and Opera extensions we have downloaded,  and also the most popular category in various datasets in recent studies~\cite{Sanc-etal-17-USENIX, Star-Niki-17-SP, Star-Niki-17-WWW}. This category does not exist on Firefox. 

We were surprised by the results that only 15 extensions (7.61\%) are classified as \code{Developer Tools}. Considering the severity of the threats, we were expecting that most of them would be extensions provided for developers to perform some controlled experiments. Since our results represent only a lower bound of the number of extensions potentially posing these risks, it would not be surprising that even more extensions also exhibit similar threats.

\begin{table}[ht]
   \centering
   \caption{Category of extensions}
   \label{tab:extensions_category}
   \rowcolors{2}{gray!25}{white}
   \begin{tabular}{|l|l|} \hline
   \rowcolor{gray!50}
      Category & \# Extensions \\
      Productivity & 81\\ 
      Social \& Communication & 48\\ 
      Fun & 19\\ 
      Accessibility & 17\\ 
      Developer Tools & 15\\ 
      Search Tools & 6\\ 
      Shopping & 4\\ 
      Blogging & 2\\ 
      Privacy \& Security & 2\\ 
      Other & 2\\ 
      Appearance & 1 \\
      \bf{Total} & \bf{197} \\ \hline
   \end{tabular}
\end{table}

{\bf{Extensions privilege only some web applications}}
About 55 extensions (45, 7 and 3 on Chrome, Firefox and Opera respectively) communicate with any web applications to give them access to extensions privileged APIs. 
Interestingly, on Chrome, 7 of them allow to execute arbitrary code in the extension context, 15 of them are concerned with SOP bypass, 26 for storing data, 2 can be exploited by any web application to read all user cookies and 5 to read the cookies of the current web application.

The vast remainder of extensions (72.08\%) can be exploited only by specific web apps to benefit from their privileged capabilities. 
For instance, reading user browsing history, bookmarks, and list of installed extensions, is enabled by extensions  only to specific applications such as \url{fliptab.io}, \url{atavi.com}, \url{mail.google.com}. In particular, downloads are allowed by many extensions (on Chrome and Opera) mostly from \url{vk.com}. 

The fact that most extensions allow communications with only some specific apps can also be explained by the fact that most of those we found allow interactions between web apps and the background pages directly. Let us recall that it is only possible to allow communications between background pages and specific web apps (and not all web apps).

{\bf{Extensions allow to connect to arbitrary web applications}}
If many extensions tend to privilege specific web applications as shown previously, the exact opposite is observed regarding the hosts extensions allow web applications to connect to, in order to access user data. For example, 37 out of the 48 extensions that can be used to bypass SOP on Chrome, give access to the user data on any other application. On Firefox, it is 6 out of the 9 extensions which allow access to any web application data. 

These two observations (extensions mostly give access to their privileged APIs only to some web applications, and allow them to access any other web application data in the case of SOP bypass) suggest that the access they give to their capabilities is rather deliberate. Moreover, for the majority of extensions, the messages to send to exploit the different APIs in extensions are so trivial that they could have only been deliberate (See Section~\ref{sec:casestudy}).

{\bf{Most privileged web applications}}
As already mentioned, most extensions allow specific apps to benefit from their privileged APIs. This is the case for instance of \url{fliptab.io} where scripts can communicate with 31 very similar HD wallpaper extensions on Chrome, that has hundreds to thousands of users. The domain \url{vk.com} can interact with 19 extensions (17 on Chrome and 2 on Opera), mostly to download files on the user device. The domain \url{atavi.com} can get access to user's history, most visited websites (topsites) and bookmarks thanks to 6 extensions.

{\bf{Extensions which pose more than one threat}}
All the extensions reported here pose at least 1 of the security and privacy threats considered in this work. Nonetheless, some extensions pose several threats.
 
The \emph{eRail.in}~\cite{eRail.in} extension on Chrome gives access to all user cookies and allows full SOP bypass from any web application. Moreover, it has more than 400k users. Interestingly, a version of the extension exists on Firefox, but it leaks cookies and data of a limited set of web applications (all related to the extension owner's domain) to the the extension's provider own domains. Five extensions provided by Fabasoft (See Table~\ref{tab:fabasoft} in the Appendix) leak the current tab cookies. As such, they allow attackers to even access HTTPOnly cookies, and use them to mount session hijacking attacks.

\emph{Ringostat dialer}~\cite{RingostatDialer} is the only extension that executes arbitrary code sent from \url{app.ringostat.com} directly in its background page. All other extensions execute the arbitrary attacker code in the context of the content scripts. Recall that the background page has access to all the capabilities an extension declares. Interestingly, it has the \host, \storage, \cookies, and \tabs\ permissions, meaning that any script present on \url{app.ringostat.com} can access user data on any other domain, access the extension storage, cookies, open new tabs, inject code directly in any tab, etc. 

\emph{StartHQ}~\cite{StartHQ} also allows to bypass SOP from \url{starthq.com}, and leaks user browsing history. Similarly, \emph{SalesforceIQ CRM}~\cite{SalesforceIQCRM}  allows to bypass SOP and leaks installed extensions to \url{mail.google.com} and \url{salesforceiq.com}.

Finally, user browsing history, bookmarks and installed extensions can be read by an attacker in  \url{atavi.com} and \url{*.fliptab.io} thanks to 6 and 31 extensions respectively (See the full list in the Appendix). The latter also let \url{fliptab.io} stores and retrieves data in the extension storage.

{\bf{Cross-browser extensions}}
It is worth mentioning that most of the extensions we found on Opera and Firefox were also present on Chrome. While the compatibility of extensions APIs on major browsers~\cite{FirefoxWebExtensionsAPI, ChromeExtensionsAPI, OperaExtensionsAPI, MicrosoftEdgeExtensionsAPI} let developers reach more users, attackers also widen their attack surface because they can impact more users thanks a single cross-browser extension. 
For instance, we have noticed that \emph{megatest2016}, an extension provider, had 2 extensions on Chrome, and a very similar one on Opera. At the time of writing this paper, Chrome removed the 2 extensions (they were allowing \url{ok.ru} and other applications to bypass SOP, but we do not know if their removal were due to the SOP bypass) while on Opera, it is still available as \emph{MegaTest - Узнать результат}~\cite{MegaTest}. 
The 2 \emph{Photo Zoom for Facebook} and \emph{Facebook Photo Zoom} Firefox add-ons have similar versions on Chrome, but these do not allow SOP bypass. Similarly, the \emph{ModernDeck} extension is present both on Opera~\cite{ModernDeckOpera} and Chrome~\cite{ModernDeckChrome}). On Opera, it allows to store/retrieve data, while on Chrome it does not. 
This represent yet another problem of cross-browser extensions. 
While users of the same extension suffer from security and privacy threats on one browser, on the other browser where the extension is removed or fixed, users do not.
Browser vendors, and more importantly users would gain from security and privacy perspectives, if browser vendors share their reviews of extensions with one another, in order to help take similar actions like removing extensions, or updating them to remove threats they pose.

\input{res_eval}

\input{res_sop}

\input{res_cookies}

\input{res_downloads}
\input{res_hbe}
\input{res_store}

\subsection{Other Threats}
\label{sec:otherthreats}
For SOP bypass, we have reported here the cases where web applications can access arbitrary data on other web applications. Nonetheless, we found many extensions allowing to access some predefined data of other web applications. This also represents a SOP bypass (since web applications cannot access such data with their normal privileges). Finally, we found some Opera and Chrome extensions (like the 31 HD wallpaper extensions by \url{fliptab.io}), and some not reported here) which allow web applications to clear user browsing data including cookies (or even set/get cookies of some specific domains), history, bookmarks, cache, stored passwords, or  enable/disable/uninstall extensions. We do not include such cases in this paper because of page limitations.

%% file: res_eval.tex
\subsection{Execute Code}
Extensions execute in browsers with elevated privileges. 
From an attacker's perspective, being able to execute arbitrary code in an extension context also gives the attacker access to the extension capabilities. 
We found 15 extensions on Chrome, 2 on Firefox and 2 on Opera that can be exploited by web apps to execute code in their privileged context. Only one extension on Chrome \emph{Ringostat dialer}~\cite{RingostatDialer} executes in its background page, code that it receives from \url{app.ringostat.com}. Then it gives access to user data on any application, user cookies, allows code injection in in any tab the user opens, the use of the extension storage, etc. All other extensions execute the attacker's code in the contexts of the content scripts. Even though content scripts  have limited access to extensions capabilities, they are not subject to SOP, can store/retrieve data, and more importantly, they have access to the full DOM on the web applications pages in which they are injected. 

The extension \emph{iwassa}, present on Opera~\cite{IwassaOpera} and Chrome~\cite{IwassaChrome} allows any app to open any URL in a new tab, and execute any code (content script) in it. If the code in the context of the content script can already access any application data, one can further inject specific content in the DOM of the new tabs opened, to exfiltrate for instance any token/secret present in the application DOM. In fact, in addition to cookies, many sensitive applications use tokens to further perform additional checks about the origins of requests before letting users perform sensitive actions on their data. 

Another interesting example is that of the \emph{LinkClicker} extension also present on Opera~\cite{LinkClickerOpera} and Chrome~\cite{LinkClickerChrome}. It allows any application to send  a code which will be further injected in any new tab the user opens during the current browsing session. One can use it to track the user while she is browsing, gather any credentials that she is providing to log into any application, and exfiltrate those to the attacker.

In many of these cases, the problem is due to the fact that the extension does not correctly sanitize the codes received from web applications, allowing attackers to execute arbitrary codes. 
%
%
A good example is that of the \emph{GureTV: To watch television} extension on Firefox~\cite{GurevertvFirefox}. It did well to sanitize content sent from web applications, but not content sent from iframes embedded in the applications. Hence, one can create an iframe, and send an arbitrary code which will be executed in the context of the content scripts. 

Many of the other extensions work similarly, and allow (at least) to access arbitrary user data on any application, and/or store and retrieve data (when they have the appropriate permissions).

%% file: res_sop.tex
\subsection{Bypass SOP}
Extensions are not subject to the SOP, and therefore have access to user data on any web application for which they have declared the \host\ permission. Through message exchanges with extensions, 48, 9 and 6 of extensions on Chrome, Firefox and Opera respectively, allow web applications to bypass SOP by accessing user data on any other web application. As for other threats, the trend is rather to allow only some web applications to bypass SOP, even though 15 of such Chrome extensions allow any application to access any other application data. Hence, the majority of arbitrary SOP bypass can be exploited by specific web applications, including: \url{ok.ru}, \url{mail.google.com}, \url{logincat.com}, etc.
Interestingly, when SOP bypass is possible, in most of the cases the data of all domains can be accessed. On Chrome for instance, it is 37 out of the 48 extensions that allow access to any application data. 
Even when the SOP bypass is partial, it is enabled to rather sensitive domains. For instance, 5 extensions out of 11 allow SOP bypass to users' Google accounts: \url{salesmate.io}, \url{appspot.com} and \url{aliexpress.com} can access users Gmail account. One extension~\cite{LinkedInSalesNavigatorChrome} allows access to the \url{linkedin.com} data of more than 400k users from Gmail, and \url{blog.renren.com} can access \url{github.com}~\cite{RenrenMarkdownChrome}.

%% file: res_cookies.tex
\subsection{Cookies}
\label{sec:results_cookies}

We found 8 Chrome extensions that can be exploited by web applications to read user cookies: 2 of them allow any web application to read all user cookies~\cite{eRail.in, TelerikChrome}, 1 only allow \code{app.ringostat.com}~\cite{RingostatDialer} to read all user cookies, and the other 5 of them allow an attacker script to read the cookies of the tab in which it executes. The number of users affected is very important (more than 415k for \emph{eRail.in}~\cite{eRail.in}, 9.6k for \emph{Telerik Test Studio Chrome Playback 2014.1}~\cite{TelerikChrome} and 78 for \emph{Ringostat dialer}~\cite{RingostatDialer}. Cookies can be used to hijack users browsing sessions, access their data and take actions on their behalf. 
It is worth mentioning that the three extensions that can be exploited to read all user cookies, have probably been poorly programmed. 
It is more likely that the ability to read cookies was meant to be used from specific web applications, but unfortunately the extensions were poorly programmed, allowing other web applications to also get access to user cookies. In particular, the \emph{Ringostat dialer}~\cite{RingostatDialer} extension did not expose any means to get user cookies. But it allows to execute any code sent from \url{app.ringostat.com} in the extension background page context (using \code{eval} function), giving the application access to all the capabilities of the extension. Among those, the cookies, storage and arbitrary host permissions, and the ability to open tabs,  inject and execute arbitrary code in them, etc.

We found that the web application \url{https://erail.in/} is effectively reading all user cookies when the \emph{eRail.in}~\cite{eRail.in} Chrome extension is installed. This means that the extension is intentionally given access to user cookies to \url{https://erail.in}.
However, it is not clear whether the cookies of interest were only those of \url{https://erail.in} or any cookie or if only cookies of \url{https://erail.in/} were meant to be leaked.
Unfortunately, any web app can access all user cookies stored by any web application, and use them to hijack user sessions. Interestingly, the extension has a version on Firefox, where the cookies which are leaked are only those of domains related to \url{erail.in} and are leaked only to \url{erail.in} and \url{eair.in}.

The case of the extension \emph{Telerik Test Studio Chrome Playback 2014.1}~\cite{TelerikChrome} is particularly interesting, as one has to setup complex interactions, involving the extension content scripts and background page, as well as the application and its server. In particular, the interactions are triggered from the web application, but the cookies are sent to the server of the application instead of being returned directly to the application. Following the same mechanism, one can clear cookies, delete user browsing history, etc. A similar extension is also available on Firefox \emph{progress-test-studio-extension}. Unfortunately, we could not analyze it as it was not downloading. 

Finally, 5 Fabasoft extensions (See more details in Table~\ref{tab:fabasoft} in the Appendix) allow the attacker to read the current tab cookies of any web application. Even though a web application protects its cookies with the HTTPOnly flag~\cite{HTTPOnly}, an attacker script running in the web application bypasses this protection by obtaining the cookies via the extension. It can further use them to mount session hijacking attacks against the user.

%% file: res_downloads.tex
\subsection{Downloads}
Exploiting extensions to trigger the download of arbitrary files is enabled mainly from specific applications including \url{vk.com} (See Table~\ref{tab:downloads_vk} in Appendix) and \url{ok.ru}. Only 2 extensions on Chrome and 3 on Firefox allow downloads from arbitrary web apps. 
The main purpose of the related extensions were to allow the download of music and videos. Sometimes, they would even suffix the downloaded file name by \code{.mp3} or \code{.mp4}. 
Nonetheless, we have been able to exploit these extensions in order to trigger the download of arbitrary files and save them in the user's device. An attacker can also do so to download malicious software, which when inadvertently executed by the user, may allow the attacker to take control of their computer and perform malicious actions.

It is worth mentioning that none of these extensions required user action to trigger the downloads. 
One of them, \emph{multiDownloader}~\cite{MultiDownloaderChrome} even overwrites a file if it is already present on the user's device. 

It is also worth mentioning the case of the Chrome \emph{repl.it download} extension~\cite{ReplitChrome}. It is a helper extension for the \url{https://repl.it} application used for creating and running programs in different languages online. The extension allows to save the code being created. Even though the extension prompts the user to confirm the file name (default is \code{program.}), the content of the file can be fully arbitrary. As such, an attacker can trick the user in saving the code being edited, while a completely different content is saved. 

%% file: res_hbe.tex
\subsection{History, Bookmarks, and List of Installed Extensions}
Two providers distinguish themselves with regards to extensions that can be exploited to get access to user browsing history, bookmarks and list of extensions. On Chrome, \code{fliptab.io}~\cite{FliptabIO} provides 31 very similar HD wallpapers extensions (See the full list in Appendix), and allows \url{fliptab.io} to get all browsing history, bookmarks and the list of user installed extensions. Each of these extensions has between a hundred and 25k users.

Furthermore, six extensions provided by \url{atavi.com} also provide the same privileges to pages at \url{atavi.com} and \url{atavi.test}. One of them, \emph{Atavi - bookmark manager
}~\cite{AtaviBookmarksChrome} has more than 96k users. 

Additionally, \emph{Browser History}~\cite{BrowserHistoryChrome} leaks user browsing history to \url{www.americaninternetmatrix.com/history}. Finally, \emph{StartHQ}~\cite{StartHQ} leaks browsing history to \url{https://starthq.com}.
Other extensions that give access to the list of extensions include \emph{Boomerang for Gmail}~\cite{BoomerangChrome} (with more than 1.5 million users) to \url{mail.google.com} and \emph{SalesforceIQ CRM}~\cite{SalesforceIQCRM}, to \url{mail.google.com} and \url{salesforceiq.com}.

%% file: res_store.tex
\subsection{Store/Retrieve Data}

About 85 extensions can be exploited by various web applications to store and retrieve data. On Chrome, 26 of these extensions give any application access to their storage. Others give specific apps access to their storage. For instance, \code{fliptab.io} can store data in the user's browser thanks to its 31 extensions. The domain  \code{netflix.com} is also able to store data thanks to 3 extensions, and \code{mail.google.com} to do so thanks to 2 extensions. The extensions \emph{ISOGG Y-Tree AddOn}~\cite{ISOGGChrome} and \emph{PhyloTreeMT AddOn}~\cite{PhyloTreeMTChrome} are from the same provider, even though the web applications they allow to persist data are respectively \code{isogg.org} and \code{phylotree.org}. 

Recall that extensions storage is persistent and not affected by the clearing of browsing data (web application cookies, storages, ...). As such, they represent a resilient storage which can be used to bypass user privacy preferences and uniquely identify them even though they have cleared their cookies. Interestingly, some extensions propose to sync data they store on all the devices the user is logged into. For instance, if a user logs into multiple devices with the same extension installed, then syncing storages lets an application tracks her accross all her devices.


%% file: casestudy.tex
\section{Case Study}
\label{sec:casestudy}
In this section, we show how an attacker can exploit the capabilities of an extension by sending the appropriate message. 
One can also find \href{http://www-sop.inria.fr/members/Doliere.Some/empoweb/extensions/}{online a few videos} demonstrating the exploits on some of concerned extensions, on the Chrome browser. 
In order to gain access to privileged browser features via an extension, an attacker first needs to ensure that the extension is installed and enabled. Many recent studies discussed extensions discovery, using for instance their unique identifiers and web accessible resources~\cite{Sjos-etal-17-CODASPY, Sanc-etal-17-USENIX} or DOM specific changes they introduce in web pages~\cite{Star-Niki-17-SP}.
%
This is not really needed here. Knowing the structure of messages extensions respond to, is sufficient. If the extension is present, it will surely reply. 
To benefit from extensions capabilities, it is sufficient that the attacker is present in a web application with which the extension can interact.

\subsection{Example of messages to send to extensions}
We refer to Section~\ref{sec:background} which presents the message passing APIs between webpages and the different components of an extension. 
Because of page limitations, we cannot provide for all extensions, the messages that can be sent from web pages to exploit extensions capabilities. We illustrate at least each threat by an extension. 

\paragraph*{\bf{Execute code in content scripts context}}
Listing~\ref{lst:execute_code_jianlibao} present the structure of messages that can be sent from any webpage to the \emph{jianlibao}~\cite{JianlibaoChrome} Chrome extension to execute arbitrary code in the context of its content scripts. Replace \code{CODE} with real JavaScript code, then serialize the message using \code{JSON.stringify} before sending it. The extension has the \storage\ and \host\ permissions meaning that any page can bypass SOP and get access to user data on any domain, store data in the extension storage and later retrieve it for tracking purposes. Moreover, the code is injected in the active tab the user is interacting with. As the user may switch tabs at any time, one can send the code regularly (say every second) in order to ensure that it is injected in all the web applications the user is interacting with. 
Since content scripts have access to the DOM of webpages, the injected code also has full access to the active tab DOM, giving it the ability to undertake any action: recording user name and password, credit card numbers, emails, etc. 

\begin{lstlisting}[label={lst:execute_code_jianlibao}, caption={Executing arbitrary code in the context of the content scripts of the current tab the user navigates to, thanks to the \emph{jianlibao} Chrome extension.}]
{
	type: "getResumeInfo", 
	downloadObj: {
        resumeWhereabouts: 5
    }, 
    context: {
        contentScript: CODE, 
        jsMethod: "console.log"
    }
}
\end{lstlisting}

Extensions such as \emph{iwassa}~\cite{IwassaChrome, IwassaOpera} or \emph{LinkClicker}~\cite{LinkClickerChrome, LinkClickerOpera}, present on Chrome and Opera, even allow to send a URL and a code. They will open the URL in a new tab, and execute the code in the context of the content scripts injected by the extension in the new tab.
Listing~\ref{lst:execute_code_iwassa} presents the case of the \emph{iwassa} extension. Replace \emph{URL} with the URL of the page to open in a new tab, and \code{CODE} with the real code to be executed in the context of the new tab content scripts. 
\begin{lstlisting}[caption={Executing code in the context of a choosen tab thanks to the \emph{iwassa} extension present on Chrome and Opera. \emph{URL} is the URL of the page to open in a new tab, and \code{CODE} the code to be executed.}, label={lst:execute_code_iwassa}]
{ 
	from: "logininfo", 
	val: [URL, CODE, "LoginAPI"]
}
\end{lstlisting} 
The extension also has the \host\ permission, allowing to make AJAX requests to any domain.

\paragraph*{\bf{Execute code in background page context}}
Background pages are the most privileged contexts, as they have access to all the capabilities of an extension. 
Listing~\ref{lst:execute_code_ringostat} shows the message to send to the \emph{Ringostat dialer}~\cite{RingostatDialer} Chrome extension to execute arbitrary code in the context of its background page. Interestingly, this extension has the \host, \storage, \cookies\ and \tabs\ permission, giving an attacker the ability to bypass SOP, store data in the extension storage, manage user cookies and tabs (open new tabs, close some, etc.). Messages are to be sent from webpages which URLs match \code{*://app.ringostat.com/*}. 

\begin{lstlisting}[caption={Message to send to \emph{Ringostat dialer} background page to execute arbitrary code. Replace \code{CODE} with the real code to be executed.}, label={lst:execute_code_ringostat}]
{
	message: "execCommand",
	data : {
		command: "eval", 
		params: CODE
  }
}
\end{lstlisting}

\paragraph*{\bf{Bypass SOP}}
Here we take the example of the \emph{Buxenger} extension, available both on Chrome and Firefox. 
Listing~\ref{lst:bypass_sop_buxenger} shows the structure of messages to be sent to the extension in order to make AJAX requests to any domain (SOP bypass). The case shown here, is for making HTTP \emph{GET} requests. But the extension also allows to make AJAX requests using HTTP \emph{POST, DELETE, PATCH} methods. 

\begin{lstlisting}[caption={Make arbitrary AJAX requests thanks to the \emph{Buxenger} extension present on Chrome and Firefox. Replace \emph{URL} with the URL of the data to access, and \emph{ID} with any value. }, label={lst:bypass_sop_buxenger}]
{ 
	message: "ajax-get", 
	url: URL, 
	callbackId: ID
}
\end{lstlisting}

\paragraph*{\bf{Retrieve cookies}}
Listing~\ref{lst:cookies_erailin} shows the case of the \emph{eRail.in} Chrome extension which allows any webpage to retrieve the list of user cookies. 
\begin{lstlisting}[caption={Message to send to \emph{erail.in} extension in order retrieve all user cookies}, label={lst:cookies_erailin}]
{
	Action: "GETCOOKIE"
}
\end{lstlisting}
This includes any cookies, such as the user authentication cookies set after she has logged into web applications. One can further use the cookies to mount session hijacking attacks. 
The extension also allows to make arbitrary AJAX requests, by sending messages as shown in Listing~\ref{lst:bypass_sop_erailin}
\begin{lstlisting}[caption={Making AJAX requests thanks to the \emph{eRail.in} Chrome extension} , label={lst:bypass_sop_erailin}]
{ 
	Action: "GET_BLOB", 
	URL: URL
}
\end{lstlisting}

\paragraph*{\bf{Downloads files}}
Listing~\ref{lst:downloads_http_commander} shows the signature of messages to send from any webpage, to the \emph{HTTP Commander}~\cite{HTTPCommanderChrome} Chrome extension in order to trigger the download of any file. Replace \emph{FILE\_URL} with the URL of the file to download, and \emph{FILE\_NAME} with the name under which the file will be saved on the user device.
Multiple files can be sent in the message. They will all be downloaded one after the other.
\begin{lstlisting}[caption={Download files on the user device, thanks to the \emph{HTTP Commander} extension.}, label={lst:downloads_http_commander}]
{   
	type: "HTCOMNET_DOWNLOAD", 
	files: [{
		url: FILE_URL, 
		path: FILE_NAME
	}]
}
\end{lstlisting}

\paragraph*{\bf{Store data in extension storage}}
Listing~\ref{lst:storage_visual} shows messages to send in order to store and retrieve data in the \emph{VisualSP Training for Office 365}~\cite{VisualSPChrome} Chrome extension storage. Replace \emph{DATA\_TO\_STORE} with the data to be stored in the extension storage. Later on, send the second message to retrieve data. The data will be sent to iframes in the page. To collect the data previously stored in the extension storage, before sending the message, one can simply add an iframe to the webpage, then send the message, collect the previously stored data from the iframe, and send it back to the parent page. 
\begin{lstlisting}[caption={Store and retrieve data in \emph{VisualSP Training for Office 365} Chrome extension storage}, label={lst:storage_visual}]
// Store data
{	
	owner: "VisualSP", 
	command: "SetUserId", 
	data: DATA_TO_STORE 
}
// Retrieve data.
{ 
	owner: "VisualSP", 
	command: "GetUserId"
}
\end{lstlisting}

\paragraph*{\bf{History, bookmarks, extensions list}}
We show here the case of the \emph{Space Galaxy HD Wallpapers}~\cite{SpaceGalaxyHDWallpapersChrome}. It is one of the 31 HD Wallpapers from \code{fliptab.io} (See Table~\ref{tab:fliptab} in the appendix) that lets pages matching \url{*.fliptab.io}, to manage user history, bookmarks, extensions list and storage. 
Listing~\ref{lst:history_bookmarks_extension_storage_fliptab} shows the different messages that has to be sent to get the related information.

\begin{lstlisting}[caption={The \emph{Space Galaxy HD Wallpapers} Chrome extension allows to get user browsing history, bookmarks and extension list}, label={lst:history_bookmarks_extension_storage_fliptab}]
// Message for retrieving user browsing history
{ 
	type: "history", 
	act: "get_all" 
}

// Message for retrieving bookmarks
{ 
	type: "bookmarks", 
	act: "get_all" 
}

// Message for retrieving the list of extensions
{ 
	type: "extensions", 
	act: "get_all"
}
\end{lstlisting}

\subsection{Forcing the attack}
\begin{figure}[ht]
\includegraphics[width=0.4\textwidth]{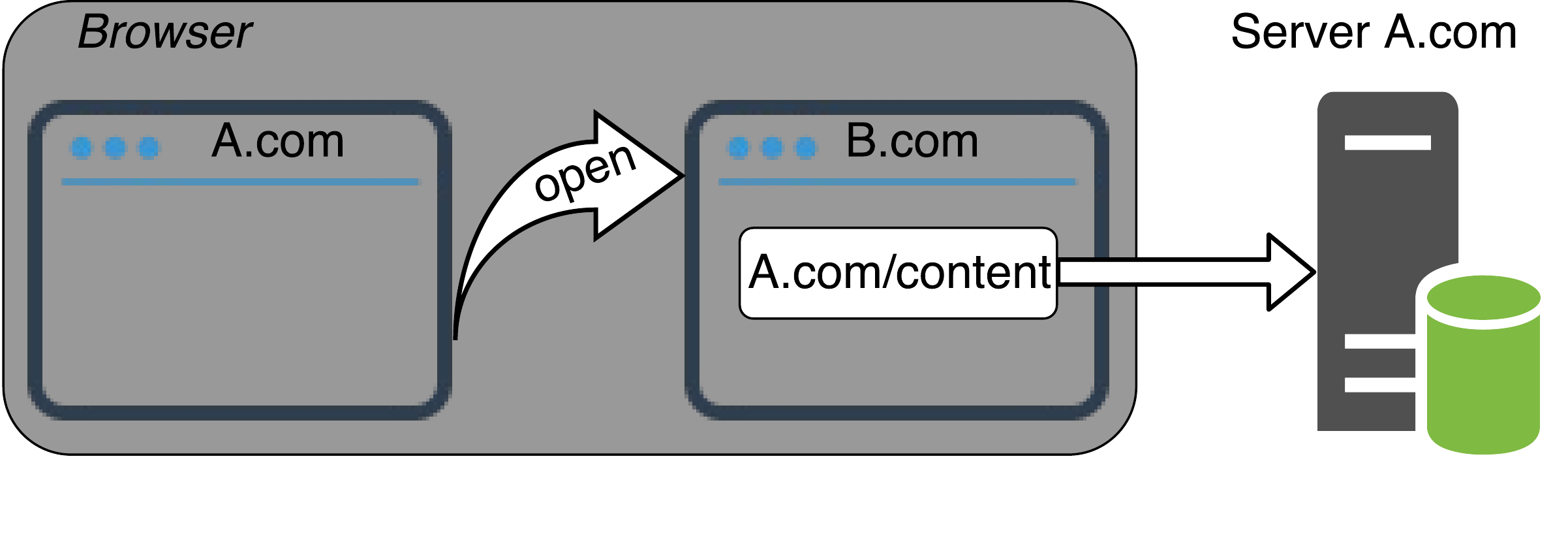}
\centering
\captionsetup{justification=centering}
\caption{\url{A.com} forces an attack by opening \url{B.com} thereby allowing \url{A.com/content} to load, execute and interact with extensions in order to exfiltrate user data to \url{A.com}.}
\label{fig:force_attack}
\end{figure}

In order for an attacker to gain access to an extension's APIs, he must have a script loaded in a web application that is allowed to interact with the extension. Moreover, in most cases, the application has to be running in the user browser in order for communications to be possible. Figure~\ref{fig:force_attack} shows a simple scenario in which \url{A.com} is an application currently running in a user browser. This application provides content \url{A.com/content} (a script) for another application \url{B.com} which can communicate with an extension to get access to some privileged APIs. However, \url{B.com} is not currently running in the user browser. \url{A.com} can force the attack to happen, by opening \url{B.com} (upon a user interaction with the \url{A.com}). Once \url{B.com} runs, the script that it embeds from \url{A.com}  gets executed and can communicate with the extension to get access to its privileged APIs --- for instance to access user data on any other application --- and exfiltrate this to \url{A.com}.
With the prevalence of some third party scripts providers among web applications~\cite{Niki-etal-12-CCS}, this scenario can be easily implemented by attackers to gain from extensions capabilities.

{\bf{Combining multiple extensions}}
Another scenario where access to any extension capabilities can be indirectly gained is when some extensions make  it possible to open new tabs and inject and execute arbitrary codes in them.
We have recorded a video showing the use of the \emph{LinkClicker} extension~\cite{LinkClickerChrome} which allows to open a new tab and execute code in it, and the \emph{Space Galaxy HD Wallpapers} extension~\cite{SpaceGalaxyHDWallpapersChrome} which allows only \code{fliptab.io} to get/delete user browsing history, bookmarks and extensions list. From any application (the \code{localhost} in our example), we opened \code{www.flipatab.io}, and injected a code in its context. The code retrieved the list of extensions, bookmarks and user history. This information could be further sent to a server chosen by the attacker. One can even use the \emph{LinkClicker} extension to send the retrieved information back to the attacker by opening a new tab of the attacker application (\code{localhost} in our case).

%% file: ndiscuss.tex
\section{Discussion}
Here we discuss countermeasures and proposals to mitigate the threats introduced by extensions via message passing.

\subsection{Disclosure to vendors}
We have disclosed the list of extensions to Chrome, Firefox and Opera. All vendors acknowledged the issues. 
Firefox has removed all the reported extensions.
%
Opera has also removed all the extensions but 2 which can be exploited to trigger downloads. 
The reason given by Opera is that the downloads can only be triggered from specific websites. However, we made them observe that those websites include third party scripts that can also trigger arbitrary downloads. 
%
So discussion still continues with Opera on the 2 remaining extensions, in particular to ensure that users are aware of the downloads. 
%
Chrome also acknowledged the problem in the reported extensions. We are still discussing with them on potential actions to take: either remove or fix the extensions. 

\subsection{Proposals}
The discussions with browser vendors confirmed our arguments that  their current extensions review process is weak. In fact, none of them has considered the fact that extensions put user data at risk via vulnerabilities in the use of message passing APIs. 
Moreover, we are worried that malicious extensions developers that would be aware of the ability to exfiltrate user data via message passing, would deliberately introduce such vulnerabilities in their extensions. There are various ways an extension could exploit its own vulnerabilities without being blocked by browser vendors. For example, the extension developer can operate a website. Then when the user opens her browser and navigates to her favorite web applications, the extension injects its own website as an iframe and exfiltrate user data from that iframe. 
For browser vendors, a quick fix of the threats discussed in this work is to consider message passing interfaces as a medium for introducing vulnerabilities in extensions, thereby putting users data at risk. New extensions must be reviewed accordingly, in order to fix such threats. To help in this process, browser vendors may mandate that extensions explicitly declare the list of web applications they intent to interact with by message passing via the extensions content scripts, very similarly to what is done with the \externally\ key used in extensions \manifest\  files to declare the list of web applications the extension background pages intent to directly interact with (See Section~\ref{sec:coms_interactions}).

The best solution to mitigate this threat would have been to ban the interactions between webpages and extensions, but this would impact the many extensions making use of these communication interfaces. Nonetheless, the needs and implementations of the message passing interfaces are questionable. 
In fact, extensions can already read/write web applications DOM. 
For extensions that absolutely need to exchange messages with webpages, browser vendors may review the current extensions system and allow messages only from code injected by extensions in the context of webpages. 
In fact, extensions can inject code directly in the context of webpages. Currently, such code runs with the same privileges as codes loaded by webpages themselves. We envision an architecture in which the browser tracks the origin of messages received in extensions. And if they are not sent by code injected by extensions, the messages is not delivered to the extension. 
There are surely ways an attacker can circumvent this solution, but such attacker is exactly the one already discussed by Carlini et al.~\cite{Carl-etal-12-USENIX} and Bandhakavi et al.~\cite{Band-etal-10-USENIX}. 

%% file: related_work.tex
\section{Related Work}

The security and privacy implications of browser extensions have been extensively studied. 
Barth et al.~\cite{Bart-etal-10-NDSS} analyzed the Firefox XPCOM architecture and proposed a new extensions architecture that has since been adopted by Google Chrome and evolved into the Chrome Extensions API compatible with the cross-browser WebExtensions API. Before them, many authors had also shown the dangers of misusing the powerful APIs provided to Firefox XPCOM extensions and propose tools for discovering vulnerabilities and securing extensions~\cite{Louw-etal-07-DIMVA, Band-etal-10-USENIX, Onar-etal-13-DIMVA, Onar-etal-15-CS}.
Among other things, the permissions system in extensions was meant to reduce extensions capabilities, and hence reduce the harms that attackers can cause if they compromise an extension. However a good number of studies have shown that many extensions still request too many permissions~\cite{Guha-etal-11-SP, Kapra-etal-14-USENIX, Heul-etal-15-HOTOS}. 
Guha et al.~\cite{Guha-etal-11-SP} proposed IBEX a cross-browser extensions platform, supporting fine-grained access control policies with tools for verifying the compliance with the security policies. The work of Carlini et al.~\cite{Carl-etal-12-USENIX} on vulnerable extensions has led to the ban of inline and HTTP scripts and eval-like functions in extensions background pages. Different dynamic analysis systems have been proposed for discovering malicious extension such as Hulk~\cite{Kapra-etal-14-USENIX} and Ex-Ray~\cite{Weis-etal-17-CSAC} based on the concept of honey pages.
Starov and Nickiforakis~\cite{Star-Niki-17-WWW} 
found many extensions leaking sensitive user information such as browsing history, search queries, form data and extensions list. 
Recently, many studies have demonstrated different techniques for discovering browser extensions and shown that they can be used to fingerprint browsers~\cite{Sjos-etal-17-CODASPY, Star-Niki-17-SP, Sanc-etal-17-USENIX, Guly-etal-18-WPES}. Other threats considered in this paper have been discussed outside of browser extensions~\cite{Roes-etal-12-NSDI, Maye-Mitc-12-SP, Engl-Nara-16-CCS, Lern-etal-16-USENIX, SessionHijacking, Johns-17-CRYPT, SessionHijacking, SOP, HTTPOnly}.

Calzavara et al.~\cite{Calz-etal-15-ESOP} were the first to show that message passing interfaces could lead to privilege escalation and exploits by web applications. We discussed directly with the authors of this work. Their goal was to formalize the privileges that an opponent can escalate thanks to the message passing APIs between web applications and extensions content scripts. In the extensions system they considered, content scripts had no privileges and direct interactions with background pages were not possible. 
They had proposed a prototype implementation of their system named CHEN for developers to evaluate the robustness of an extension against privilege escalation and help them refactor their codes, but the tool is no more available and it did not take into account long-term communications (ports)

To the best of our knowledge, this work is the first large-scale study on the security and privacy implications of the communications between browser extensions and web applications, allowing the latter to benefit from extensions privileged capabilities. We built a static analyzer for analyzing extensions and identified a good number of them, demonstrating how these extensions can be exploited by web applications to benefit from extensions privileged capabilities and thereby access sensitive user information. 

%% file: conclusion.tex
\section{Conclusion}
Browser extensions are third party code in browsers with access to privileged APIs not accessible to web applications. Nevertheless, web applications and browser extensions can interact with one another by exchanging messages. 
In this paper, we built a static analyzer and applied it to Chrome, Firefox and Opera extensions. We identified a good number of extensions that can be exploited by web applications to benefit from their privileged capabilities. In particular, some vulnerable extensions allow  web applications 
to bypass the Same Origin Policy security mechanism and access user data on any web application. Extensions also leaked user credentials (cookies), browsing history, bookmarks, list of installed extensions, to web applications or allowed them to download any file on the user device, or store data in the extension storage for tracking purposes. 
We showed how trivially, attackers can exploit those threats, and discussed proposals as to mitigate them. In particular we argued for a review process taking into consideration the threats we have discussed, with the help of tools such as our static analyzer, or changes in the extensions system itself to ban or limit messages only to extension injected scripts.

%% file: extensions.tex
\begin{table}[h]
\caption{Extensions with the same code base which gives *.fliptab.io access  to browsing history (get/delete), bookmarks (get), extensions (get/enable/disable/uninstall) and storage}
\label{tab:fliptab}
\begin{tabular}{l}
bddmmehmgpjhhmbbmngdjhlednmkbken, 
cajmbfbhhfelhgolhldhhodkclpakcfe \\ 
cepmfckfppjpbkjgnpokojedlngflnca,
clkodoejadlbjaopcjoijihebbgipjff \\ 
dekpebffaadijeaogggfhjemdbjgbcao, 
dkpndikhfepllbpaafgcelembimabofo \\ 
eeiedbnahjonkmimigblgchlefcklhok,
efdddbobcofamdjmekphjlhgmcnhobbp \\ 
ehmhopjniedignnkdeijmpmodhcppgif,
eilbnnflfpkhhfmhmlhflhecceajpkcj \\ 
fieoemdbopiialnojhifcndkenhjkbmm,
fkpmpnljocdllgmplhnmjhjmmilbnofj \\ 
gfgchcclfmppnfoakdlhgdhnolbpiedf,
glfbbjdfmmlanpikdedpjoeimlijjcjj \\ 
hmbedbiicehadpbhbipafffieolpjolh,
hocncjdhccalpmblkpagbmjebkfkibbm \\ 
iamlligjelallbdddajmbojjjhadkmcf,
jcffnpjkbahanenhcnhhdfopkjlpflfm \\ 
jokpapkhjeahjbkemfjfhjgcogmbcpoi,
kkejopfphkmldfpdmcljfoinfcljijjf \\ 
klfeojnepdoehgddffbcjiamcjjahmgj,
lbfidebeingoondbmpeapjoeeoloanak \\ 
lgphbplfjpemcghfcoajehcmikflcbbd,
lmbcpiodajlbgmjbiajgcjdalgbofcbn \\ 
loggojfoonblkkhkjpijapeheoogagki,
lpkfidfkgflpbakdnhpojiejlpdanknh \\ 
mgmodhbknbfmpjmilankiffnjbelcipo,
mibaeahdcconphmdndbeipegldkkbcjh \\ 
odpiaedkmdpcheddbkilnkelhhocoenn,
pfdaccgdljiifplhfnjcacapfedngonb \\ 
afddmpnodjaifgjibafjcbfaplnoipei \\
\end{tabular}
\end{table}

\begin{table}[h]
\caption{Extensions with the same code base for triggering downloads from vk.com, *.vimeo.com, *.coub.com, *.kinopoisk.ru }
\label{tab:downloads_vk}
\begin{tabular}{l}
nfhipbkhabgmkhahoaagkcgppcjikjgl  
idenapkfefkbknhbmfgeaclpcpbhcnbe  \\
fnnlocjimhjpmgfjhjamdkjhemfhkhjo  
lmlnplkfbiihcpkghkkmfefjdaccmbcc  \\
kbiocjbkoohjjkkeaafiemjeidgalllh  
dccmnjciogmmahaogjgkocongokmieog  \\
ekfkljjojhnnhfedepfnbhhfjklagngk  
hhfgpbjpilbbaomjmdpnfchbpipehiif  \\
pgajmafmbajahclonccaoaoleghhnpam  
ipeeopcjpgcbgnfogjlickeilmkbonen  \\
jfpmehlefcchhhmlmennihbbihaolabk  
kcollknpphnodcjdkcmgpjmlbaenabao  \\
backekeabechifnekobfachchocbmjag  
mfpbgndgoogfplejodpbhnfmaibnalkf  \\
ojhheobonaamlhlcdngacakdcigpeokl  
mienmjdbnnpaigifneeiifdbjkdgelha  \\
amaobfendgcolppeioeageanmillkmkc  \\ \hline
\end{tabular}
\end{table}

\begin{table}[h]
\caption{Extensions with the same code base which leaks topsites, history and/or bookmarks to *.atavi.com, *.atavi.test}
\label{tab:atavi}
\begin{tabular}{l}
iglbnbabjdfaobglhonmnlkdbommiebd,   
knflcnelciofoghldagpknelepafjeif  \\ 
lamnafpjcnoclihgpefhdbefcmjikhaj,   
jffjjdoccjiflmckicphblggbppfgklk  \\ 
ofmacdiceehcibkfednmgpkhgfhpacgi,   
jpchabeoojaflbaajmjhfcfiknckabpo \\  

\end{tabular}
\end{table}

\begin{table}[h]
\caption{Extensions with the same code base, provided by Fabasoft, which give access to the current tab cookies}
\label{tab:fabasoft}
\begin{tabular}{l}
ajlbdflhaaflcepndpkdgejimggjcpnm,
ngbcdblbfdpjgpmgfagkfofcjbnggfgn \\
pdhjoolhbkmlgjfedckdhiknnoabbnkk,
hiejidhjgjpelfgldfhmnaoahnephhfg \\
icjlkccflchmagmkfidekficomdnlcig \\
\end{tabular}
\end{table}

\begin{table}[h]
\caption{Extensions which give access to their storage to any application}
\label{tab:storage_all}
\begin{tabular}{l}
eljhpoopiapggnlfcilpbihgbgbpnkgd,
akhamklknibionleflabebgeikdookmp \\
hebabhddakflgmlhgefakkfkciijliie,
ilgdjidfijkaengnhpeoneiagigajhco \\
ohdihpdgfenligmhnmldmiabdhflokkh,
abenhehmjmoifipfpjeaejpbeeihnokp \\
ackpndpapmikcoklmcbigfgkiemohddk, 
ceogcehidijhepckebfifkpfogkajdkg \\
cgijoonmpaboophnagdckdcekmpfokel, 
dhcfokhhmhenbfmeflifppiedabfggkj \\
dhcmolikocplmafolinkncghmahimooh,
eamjolanjdmgochipodfokkfjaeifhon \\
efhbachoakbcmbcmfffdgphbpcbldjac,
fecipnolpdcmoidbjbnakpjgfikbnaik \\
gnnagpehbmfalanfjadamobejlldgedo,
ijdfpccaiklfhpnamolipbjjijilmhli \\
khjhfgcimhcnaimdbgjbnbhcojkoceoc,
niceocbendibobemckcagggppphheomc \\
okcfiidnmioajibmhhjpiomgejajiafa,
pjjceionkajpednnegoanjjdlhbgkkpc \\
pjojmkmdealampgchopkfbejihpimjia,
\end{tabular}
\end{table}

\begin{table}[h]
\caption{Extensions which give access to their storage to specific applications}
\label{tab:storage_some}
\begin{tabular}{ll}
lpkhcobfjeidpkllbeagkkmmjgbmpfch & mail.google.com \\
eggdmhdpffgikgakkfojgiledkekfdce & mail.google.com \\
jmllflbhbembffempimjdbgnaodpoihh & mail.google.com \\
jmlnhlclbpfcbkaoaegfigepaffoankc & *.google.com, \\
gaoiiiehelhpkmpkolndijhiogfholcc & netflix.com \\
ghldlmcbffbcnoofadgcapodmpiimflj & netflix.com \\
jpgadigdffhcjldfkanacncocacekkie & netflix.com/watch/ \\
peiajekggpiihnhphljoikpjeaahkdcn & beam.pro \\
bnfboihohdckgijdkplinpflifbbfmhm & plug.dj \\
aclhfmpoahihmhhacaekgcbjaeojnifa & wordix.io, \\
hcdfoeppbchkbbpplllggbjkkfokifej & *.vk.com/feed/ \\
hddnlanhlmifafibmlabomkkkobcmchj & thankscoin.org \\
lhjajgnfmiliphkioedlmbfcdkhdhnkc & *.service-now.com \\
bmdlalnebjigindhobniianfmhakfelf & robertsspaceindustries.com, \\
dadggmdmhmfkpglkfpkjdmlendbkehoh & openvideo.droppages.com \\
pbpfgdgddpnbjcbpofmdanfbbigocklj & tweetdeck-enhancer \\
ilpkhojfiejdbkgcjbmllngjebdoehim & *.phylotree.org \\
cfnjeahambijfdljfacldifapdcklhnj & isogg.org \\
cjkbjhfhpbmnphgbppkbcidpmmbhaifa & *.player.me \\
ddiaadobgihkgefcaajmkjgmnjakiamn & auth.digitalkeyway.com \\
dienbdhbgkpddlgaceopelifcjpmkeha & *.gestionderesidencias.es \\
dnpdkejhfeeipmklhlkdjaoakbkjkkjn & datalane.io \\
gmjdaaahidcimfaipifeoekglllgdllb & chat.stackexchange.com \\
kfodnoaejimmmphonklghkimhnhhgbce & overlayBI.com \\
\end{tabular}
\end{table}

\clearpage
\onecolumn
\begin{savenotes}

\small{
\begin{longtable}{lp{3cm}p{2.5cm}p{6cm}}

\caption{Chrome, Firefox and Opera extensions which give web applications access to privileged APIs} \\
\label{tab:results_all} \\
\hline Extension unique identifier or name & Web applications to send messages from & Target web applications to access & \bf{Permissions (accessible privileged API)} \\ \hline

\multicolumn{4}{c}{\it{Chrome Browser}} \\ \hline
\endhead

fimckmjeammfdcpldmcigeojkkmeeian & * & * & \eval, \host, \storage, \downloads \\
fidaihkgnbcbkkdaoebdionfjenegede & * & * & \eval, \host, \storage \\
hnkmipajjgbclkombnmigfnpekddlhlh & * & * & \eval, \host, \storage \\
fajjnmbcianlnhmngmabhgkmgdindlha & * & * & \eval, \host, \storage \\ 
efajnkcfjjkcodbhkhaigkffdleomnag & * & * & \eval, \host \\ 
hoobpdoclliidciecjifpikpnopjpmkh & * & * & \eval, \host \\ 
kjfjdocojijlledbaanbhpcnkoimghal & * & * & \eval, \host \\ 
pfofjhnkanlacmgfgjohncmgemffkldl & app.ringostat.com & * & \eval, \host, \cookies, \storage \\ 
gooecknlakggnppmhfpopneedjconjjp & lionlock.com, & * & \eval, \host, \storage \\ 
bdiogkcdmlehdjfandmfaibbkkaicppk & *.delfa.com.br & * & \eval, \host, \storage \\ 
pgbjjemkcflenaakhiehfdmcdnlnlpbl & www.seejay.cloud & * & \eval, \host, \storage \\ 
hdanmfijddamndfaabibmcafmnhhmebi & *.hirogete.com, & * & \eval, \host \\ 
hpmeebiiihmjelpjmmemlihhcacflflc & *.valleyge.com & * & \eval, \host \\  
oejnkhmeilmiplpmenkegjaibnjbappo & search.lilo.org, & * & \eval, \host \\ 
jkoegdibpkleifbkojmplebjhfllkckn & search.uselilo.org, & * & \eval, \host \\ 

aopfgjfeiimeioiajeknfidlljpoebgc & * & * & \host, \cookies \\ 
hlagecmhpppmpfdifmigdglnhcpnohib & * & * & \host \\ 
kpgdinlfgnkbfkmffilkgmeahphehegk & * & * & \host \\ 
bjjpnhdlhpfdebcbhdlmecafnokpjpce & * & * & \host \\ 
bmiedopcajpcehbbfglefijfmmndcaoa & * & * & \host \\ 
jegnjmcegcpodciadcoeneecmkiccfgi & * & * & \host \\ 
jnhibbjmekoijdjaopflcjbjieamifhh & * & * & \host \\ 
jpkfmllgncphdgojhkbcjidgeabaible & * & * & \host \\ 
ilcpdgfepihaomggobhmfiimflngbcoh & starthq.com, & * & \host, \history \\ 
jpcebpeheognnbogfkpllmmdnimjffdb & mail.google.com, & * & \host, \management \\ 
cnkgdfnjmgamkcpjdljdncfjcegpgcdg & mail.google.com, & * & \host, \management \\ 
cfddhmlokgokhcmepddjooekhmgmgfld & *.ok.ru & * & \host, \downloads \\ 
efhgmgomhamkkmjbgmcpgjnabcfpnaek & *.ok.ru & * & \host, \downloads \\ 
djhfcchmdelggndcpkgbanfhnpbbijdb & *.ok.ru & * & \host, \downloads \\ 
fhlkioimlijffnblckmdikkadobdmlgn & *.apistop.com & * & \host \\ 
angncidddapgcmohkdmhidfleomhmfgi & logincat.com, & * & \host \\ 
lndhlcaobijohmgoikmgpgbhepkbhpkl & oneom.tk & * & \host \\ 
olpheomfiimdonpboopcailehdagfhaa & .g3user.com, & * & \host \\ 
idkghekmllmjgnmbohakcddgcclanlca & ln.io & * & \host \\ 
mhdhcccejcjfanablmohbpdbepdkokkj & *.gvt.com.br, & * & \host \\ 
plfffminkgohddbooidppccppgelajfp & mp.weixin.qq.com, & * & \host \\ 
cboekbiaoabkhgjdclenjpipclabkdga & *.apiary.io, & * & \host \\ 
ekeefjfdbaakgbfbagacmckiedkmakem & *.salesmate.io, & mail.google.com, & \host \\ 
lbjbbkhljiimahdeknpckaoiinopofhl & *.appspot.com & mail.google.com, & \host \\ 
ijmbknjhacbaeeoamjajoolgjgdbpkko & *.aliexpress.com, & *.google.com, & \host \\ 
hihakjfhbmlmjdnnhegiciffjplmdhin & mail.google.com & linkedin.com,  & \host \\ 
cfbodcmobhpfbjhbennacnanbmpbcfkd & *.aliexpress.com, & appfreaker.com & \host \\ 
ommfijfafanajffiijecdlfjlbgpmgpl & *.treesnetwork.com, & docs.google.com, & \host \\ 
okgfglgogpkomipfflpajohdkaflndoh & ouramazinghome.com & www.google.com & \host \\ 
iiabjaofopjooifoclbpdmffjlgbplod & blog.renren.com & *.github.com & \host \\ 
mcdjehgaflnlmilhefigdkldfdnembhk & *.spotsetter.com & *.amazonaws.com & \host \\ 
lfekjajdgncmkajdpiadkkhhpblngnlc & sub.watch, & zooqle.com, & \host \\ 
gkfpnohhmkonpkkpdbebccbgnajfgpjp & squares.io/fetch, & www.nytimes.com & \host \\

pkkbbimilpjmghfhhppamgigileopnkc & * & * & \cookies \\ 
5 Fabasoft extensions (See Table~\ref{tab:fabasoft}) & * & current tab & \cookies \\ 

emiplbkkiabideffmpogkbbogkmofgph & * & - & \downloads \\
17 extensions (See Table~\ref{tab:downloads_vk}) & vk.com, & - & \downloads \\

eadbjnlpeabhbllkljhifinhfelhimha & ok.ru & - & \downloads \\

ngegklmoecgejlbkiieccocmpmpmfhim & *.tribecube.com & - & \downloads \\
iogibhaacmieogkdgebfbjgoofdlcmgb & *.shutterstock.com & - & \downloads \\
ooeealgadmhdnhebkhhbbcmckehpomcj & animevost.org & - & \downloads \\
dnohbnpecjinmdpeikpnmheeepnapfci & vtop.vit.ac.in & - & \downloads \\
pgmcojeijjhacgkkjaakdafmloncpema & repl.it & - & \downloads \\
hacopcfnbokiahlppemnlneooamldola & hypem.com & - & \downloads \\

bpkphnbpiagbpinglgejckickdgaghjo & amer...matrix.com & - & \history  \\ 
fheihcbdclkdoeadmjfggiamjgkippli & .my-lucky-star.net & - & \topsites  \\ 
llelondjpcjljnjihdflhpclcpbiaiba & *.msn.com & - & \topsites  \\ 

6 Atavi Extensions (See Table~\ref{tab:atavi}) & atavi.com, & - & \history, \bookmarks, \topsites \\ 

31 HD Wallpapers (See Table~\ref{tab:fliptab}) & fliptab.io & - & \history, \bookmarks, \management, \storage \\

pnbfclligibfgdknphcodpbcejnkhffp & * & - & \bookmarks \\ 

eihbcgffjehfcgafjljohecmadcefoji & app.launch.menu & - & \bookmarks  \\ 
empgohlokhdhhchkenknobacofijiffg & app.launch.menu & - & \bookmarks  \\ 
aefmgkhgcmdljpfijlohmbhkhflmbmfi & openoox.com & - & \bookmarks \\ 
dhjhphjhpcelebeagllljbfpipdfkhgi & .azurewebsites.net & - & \bookmarks  \\
jeabbgpkliknjiacfkfglknajloappkh & yeahap.com, & - & \bookmarks  \\ 

22 Extensions (See Table~\ref{tab:storage_all}) & *  & - & \storage \\
24 Extensions (See Table~\ref{tab:storage_some}) & mail.google.com, ... & - &  \storage \\ \hline
\multicolumn{4}{c}{\it{Firefox Browser}} \\ \hline

guretv-ver-tv & * & * & \eval, \host, \storage \\ 
buxenger & * & * & \eval, \host \\
  
bitbucket-server & * & * & \host \\ 
logincataddon & logincat.com, & * & \host \\ 
facebook-photo-zoom-easy & www.facebook.com & * & \host \\ 
facebook-photo-zoom & www.facebook.com & * & \host \\ 
markanabak-eklentisi & *.markanabak.com, & *.wipo.int, & \host \\ 
skimdaddy & * & skimdaddy.com & \host \\ 
the-trees-network & *.treesnetwork.com, & docs.google.com, & \host \\ 

assina-me & * & - & \downloads \\
liber-capital & * & - & \downloads \\
video-downloader-1 & * & - & \downloads \\
openvost & animevost.org & - & \downloads \\
youtube-video-download-convert & *.youtube.com & - & \downloads \\

openvideo & droppages.com & - & \storage \\ 
vgis & *.vonage.com & - & \storage \\  \hline

 \multicolumn{4}{c}{\texttt{Opera Browser}} \\ \hline
bmjcngclkmgpfbjcmnbidognkoocpllm & * & * & \eval, \host, \storage \\ 
jnmcfakfglphcmgokeeoihifcenjjcgg & * & * & \eval, \host \\  

pmpnemphhmmpkcafgpdjanghiaadfbef & *.ok.ru & * & \host \\ 
mpaghnpkgmnikepcgjddhckcedapomkp & *.ok.ru, *.vk.com, & * & \host \\ 
bcabkcaakkjfdlodkolfagbdejhhkigp & *.lazyrobin.ru & *  & \host \\ 
bidjmocompdljmeglljcoecikgogfjbb & sub.watch, & zooqle.com, & \host \\ 

aghgmcnoiflhcnfjkckofmjbeinjkena & vk.com, & - & \downloads \\
mhjbdafcpnoapkglmldoofhhbpnogehk & vk.com, & - & \downloads \\

hajlecmoacenahambneialopbpleihjn & * & - & \storage \\ 
lkdpdiepahdagdknbbjgnadholcdgfib & tweetdeck-enhancer & - &  \storage \\

\end{longtable}
}
\end{savenotes}
\twocolumn